\documentclass[10pt]{iopart}

\usepackage{iopams} 
\usepackage{caption}
\usepackage{subcaption}

\usepackage{graphicx}
\usepackage[dvipsnames]{xcolor} %[expansion]
\usepackage{caption}

\usepackage{booktabs}
\usepackage[numbers,sort&compress]{natbib}
\usepackage{url,hyperref}
\hypersetup{colorlinks=true,
linkcolor=blue,
citecolor=purple,
filecolor=purple,
urlcolor=blue}

\usepackage{setspace} %paragraph/line spacing 
\AtBeginDocument{\setstretch{1.125}}
\usepackage{outlines} %nested lists! %\1 \2 \3
\usepackage[splitrule]{footmisc} %delineate footnotes

\usepackage{tensor} %neat indices
\usepackage{array}

\usepackage{slashed}
\usepackage{cancel}
\usepackage{verbatim}
\usepackage{csquotes}

\newcolumntype{C}{>{$}c<{$}} % math-mode version of "l" column type
\newcolumntype{L}{>{$}l<{$}} % math-mode version of "l" column type
\usepackage{multirow}

\begin{document}

\title[Quasinormal frequencies in Reissner-Nordstr{\"o}m de Sitter black holes]{Quasinormal frequencies in Reissner-Nordstr{\"o}m de Sitter black holes: constraints from space-time and scalar field parameters 
}

\author{Anna Chrysostomou$^{1}$, Alan S. Cornell$^{2}$, Aldo Deandrea$^{3,2}$, and Seong Chan Park$^4$}
\address{$^{1}$Laboratoire de Physique Th\'eorique et Hautes \'Energies - LPTHE, Sorbonne Universit\'e, CNRS, 4 Place Jussieu, 75005 Paris, France
}
\address{$^{2}$Department of Physics, University of Johannesburg, PO Box 524, Auckland Park 2006, South Africa}
\address{$^{3}$Universit{\'e} Claude Bernard Lyon 1, IP2I, UMR 5822, CNRS/IN2P3, 4 rue Enrico Fermi, 69622 Villeurbanne Cedex, France}

\address{$^4$Department of Physics and IPAP, Yonsei University, Seoul 03722, Republic of Korea}

\eads{\mailto{chrysostomou@lpthe.jussieu.fr}, acornell@uj.ac.za, 
deandrea@ip2i.in2p3.fr, sc.park@yonsei.ac.kr}

\vspace{10pt}
%\begin{indented}
%\item[]\today 
%\end{indented}

\begin{abstract}
We examine the quasinormal modes exhibited by a massive scalar test field carrying an electric charge, oscillating in the outer region of a Reissner-Nordstr\"om de Sitter black hole. We examine the quasinormal modes effective potential throughout the black hole mass-charge phase space, finding a single-peaked barrier potential on $r_+ < r < r_c$ for all non-extremised black hole solutions for $\ell \geq 1$. Unlike in the Schwarzschild background, increasing scalar field mass heightens the peak of this barrier potential, while increasing the scalar field charge suppresses it. We compute the corresponding quasinormal frequency spectrum using a WKB-based semi-classical method, where, like the Schwarzschild case, we observe anomalous QNM damping behaviour for small scalar field mass below some critical mass $\mu_{crit}$, which is $\ell$-independent for $q=0$. 
\end{abstract}

%%%%%
%INTRODUCTION
%%%%%

\renewcommand{\theequation}{\arabic{section}.\arabic{equation}}
\section{Introduction \label{chap:intro}}

\par When studying the quasinormal modes (QNMs) of a test field oscillating in the background of a charged black hole embedded in de Sitter space-time (i.e. the Reissner-Nordstr\"om de Sitter (RNdS) black hole solution), there are a number of free parameters that must be taken into account. In this work, we will classify these as \enquote{space-time} and \enquote{field} parameters: 

\par The \enquote{space-time parameters} are the black hole mass $(m_{\rm BH})$ and charge $(q_{\rm BH})$ that are intrinsic to the black hole per the \enquote{no-hair conjecture} \cite{Israel1968_NoHair-RN}, as well as the cosmological constant $(\Lambda)$. Recall that $\Lambda$ is a model-dependent degree of freedom belonging to the \enquote{space of theories}, where for inflationary cosmology $\Lambda>0$, and is associated with the vacuum energy of the scalar field that drove inflation in the early universe \cite{Bousso1996_Nariai}.

\par The \enquote{field parameters} include the spin of the field $(s)$ and the integer \enquote{overtone number} $(n)$ that monotonically labels the QNMs, serving as the \enquote{order} of the mode. There are a countable infinity of overtones, beginning  with the \enquote{fundamental mode} of $n=0$. The inclusion of additional field parameters is implicitly dependent on the symmetries of the space-time. For example, in the case of the spherically symmetric black holes considered here, the field parameters are limited to the integer \enquote{multipolar}/\enquote{angular momentum} number $(\ell \geq 0)$ related to the spherical harmonic decomposition of the QNMs (see Section \ref{chap:formalism}). For the $\phi^4$ scalar field theory that we study here, we must also take into account the scalar field mass $(\mu \geq 0)$ and scalar field charge $(q \geq 0)$. Unless explicitly stated otherwise, we employ Planck units $(c= \hbar = k_B =1)$. Under these units, $\mu$ has units of inverse length and $q$ is dimensionless. The Planck mass is then expressed solely in terms of Newton's gravitational constant, following the conventions of Refs. \cite{vanRiet2019_FLevapBHdS,vanRiet2021_FL}, $M^2_P = (8 \pi G)^{-1}$. 

\par The computation of massive scalar quasinormal frequencies (QNFs) within Schwarzschild black holes and beyond has been studied extensively (see Refs. \cite{refBertiCardoso,refKonoplyaZhidenkoReview} and references therein). From examples such as Refs. \cite{SimoneWill1991_MassiveScalar, Ohashi2004_MassiveScalarQ,Konoplya2004_MassiveScalar,Dolan2007_MassiveScalarKerr,refDecanini2011}, we have a general understanding of the influence of scalar field mass on the QNF spectrum. That is, for a fixed $\ell$ and small values of $\mu$, the oscillation frequency increases with increasing $\mu$ and the damping decreases to zero. At this point the QNMs enter the \enquote{quasi-resonance regime}, within which standard methods of calculating QNFs are not necessarily reliable. For example, the WKB method roughly fails when $\mu^2$ exceeds the height of the QNM potential barrier, as discussed in Ref. \cite{Konoplya2019_recipes}. For increasing multipolar numbers and fixed $\mu$, we expect the QNF and the magnitude of its damping to increase with $\ell$. In Ref. \cite{Lagos2020_Anomalous}, however, \enquote{anomalous} behaviour was observed for very small values of $\mu$ below some critical mass $(\mu_{crit})$, where the magnitude of the damping \textit{decreases} with increasing values of $\ell$. Such anomalous behaviour has been observed in Kerr space-times \cite{Lagos2020_Anomalous}, as well as in RN and RNdS space-times \cite{Fontana2020_RNanomalous,Papantonopoulos2022}. 

\par In this work, we explore massive scalar fields charged under a $U(1)$ gauge symmetry, and explore how their QNFs evolve for different charged black hole solutions in asymptotically-de Sitter space-time. Our interest in the RNdS space-time is motivated by the influence of the cosmological constant on the phase space of the charged black hole: the mass of the black hole is bounded from above upon introducing $\Lambda>0$.

More subtle concerns include the impact of the Gibbons-Hawking entropy \cite{GibbonsHawking1977_BHTherm} on the black hole system, particularly with respect to black hole decay (see Ref. \cite{MossNaritaka2021_KerrdSevaporation} and references therein). In the RNdS case, black hole emission serves as a transition towards thermal equilibrium mediated by an exchange of mass and charge between the black hole and the cosmological horizon. This process was recently investigated in Refs. \cite{vanRiet2019_FLevapBHdS,vanRiet2021_FL} to understand how RNdS black holes decay and to extend the principles of the \enquote{Weak Gravity Conjecture} \cite{ArkaniHamed2006_WGCorigins} to de Sitter space-times. This led to the \enquote{Festina-Lente} bound for elementary particles of mass $m$ and charge $q$ being discharged from the black hole: 
\begin{equation} \label{eq:FLbound}
m^2 \geq \sqrt{6} g_1 q M_P H \; ,
\end{equation}
\noindent where $g_1$ is the $U(1)$ gauge coupling and $H^2=\Lambda/3$. In Planck units, the Planck scale is $M_P \sim 10^{27}$ eV and the current Hubble scale is $H \sim 10^{-33}$ eV. All Standard Model particles satisfy the Festina-Lente bound \cite{vanRiet2019_FLevapBHdS,vanRiet2021_FL}. Moreover, since the mass of the electron is determined by the vacuum expectation value of the Higgs, the Festina-Lente bound can also be used to constrain the shape of the Higgs potential, as discussed in Ref.
\cite{Lee:2021cor}. Whether the Festina-Lente bound holds beyond the Standard Model is an active area of investigation (see, for example, Refs. \cite{SCParkDYCheong2022_FLmili,Guidetti:2022AFL}). %For the lightest electrically charged particle in the Standard Model, the electron, the particle mass $m_e \sim 10^5$ eV comfortably satisfies the Festina-Lente bound.

\par The Weak Gravity Conjecture and the Festina-Lente bound emphasise the relevance of black hole mechanics in constraining effective field theories. Within the QNM context, the nature of the space-time is explicitly considered in the boundary conditions of the eigenvalue problem, as we shall demonstrate in Section \ref{chap:RNdSQNMs}. Since mass and charge are free parameters in a scalar field theory, without \textit{a priori} motivations for their magnitudes, it is not immediately obvious how to impose reasonable constraints on their values within QNM contexts. For this reason, we focus as a first step on understanding the influence of scalar field mass and charge for QNFs within the RNdS phase space, extracting quantitative bounds where possible on the QNM behaviour from the extremised black hole background.

\par The paper is laid out as follows: we begin with a review of the RNdS black hole in Section \ref{sec:RNdSBH}, where we discuss the $(M,Q)$ phase space (Section \ref{subsec:RNdSphasespace}) and the corresponding thermodynamics (Section \ref{subsec:RNdSphasespace2}). We then review the formalism for the QNM problem in Section \ref{sec:QNMsRNdS}, focusing on the $\ell = 0$ case in Section \ref{subsec:super} and the superradiant instabilities that are observed only for vanishing multipolar number. In Section \ref{sec:potRNdS} we study the potential for $\ell \geq 1$, where we discuss our observations of the qualitative influence of space-time and scalar field parameters on the QNM effective potential. We perform our study of the QNF spectrum in Section \ref{sec:QNFspecRNdS}, where we first specify the semi-classical method that we use in this work (Section \ref{subsec:super}), consolidate QNF classifications in Section \ref{subsubsec:classQNM}, and then focus on the anomalous QNF behaviour and constraints thereof in Section \ref{sec:critmass}. We close this discussion with a comment in Section \ref{sec:ConnectionGW} on the implications of the Festina-Lente bound on the massive, charged QNF spectrum. Finally, we summarise and discuss our findings in the final section.

%%%%%
%SECTION II
%%%%%

\setcounter{equation}{0}
\renewcommand{\theequation}{\arabic{section}.\arabic{subsection}.\arabic{equation}}
\section{Einstein-Hilbert-Maxwell theory in de Sitter space-time \label{chap:formalism}}

\par Since QNMs are beholden to physically motivated boundary conditions imposed on the black hole space-time, a QNM study must be preceded by an analysis of the fixed background upon which these perturbations propagate. Here, we provide a short overview of the RNdS black hole solution and its corresponding black hole thermodynamics. The objectives of this section are to establish the phase space for our QNM study, by outlining the constraints on black hole mass and charge for a fixed value of $\Lambda$, and to provide the necessary formalism needed to discuss the QNMs of the RNdS black hole. This overview is informed by Refs. \cite{Romans1991_ColdLukewarmRNdS,Belgiorno2009_chargedBHs,Belgiorno2010_chargedBHs,AntoniadisBenakli2020_WGCdS}.

\subsection{The Reissner-Nordstr{\"o}m de Sitter black hole \label{sec:RNdSBH}}
\par Consider the Einstein-Hilbert-Maxwell action in an asymptotically-de Sitter space-time under natural units,
\begin{equation} \label{eq:action}
S =  \int d^4x \sqrt{-g} \left[ \frac{1}{2 \kappa^2}\left(R - 2\Lambda  \right) - \frac{1}{4g_1^2} F_{\mu \nu} F^{\mu \nu} \right] \;.
\end{equation}  
\noindent Here, $\kappa^2 = 8 \pi G = M^{-2}_P$ relates the gravitational coupling $\kappa$ to Newton's gravitational constant $G$ and the Planck mass $M_P$. The relationship between the cosmological constant $\Lambda >0$, the de Sitter radius $L_{dS}$, and the Hubble parameter $H$ can be expressed explicitly as $H^2 = L^{-2}_{dS} = \Lambda /3$. The Ricci scalar curvature $R=g^{\mu \nu} R_{\mu \nu}$ and the metric $g=\det |g_{\mu \nu}|$ encode the geometry of the space-time. The metric tensor $g_{\mu \nu}$ depends on the characteristic black hole parameters, $(m_{_{\rm BH}}$,$q_{_{\rm BH}}, \Lambda)$. For the electromagnetic field strength tensor $F^{\mu \nu}$, $g_1$ is the $U(1)$ gauge coupling. Since our focus is on the electrically charged RNdS black hole, the non-zero component of $F^{\mu \nu}$ is
\begin{equation}
F_{tr} = \frac{g_1^2}{4 \pi} \frac{q_{_{_{\rm BH}}}}{r} 
%dt \wedge dr
\;,
\end{equation}
with a purely electric gauge potential,
\begin{equation} \label{eq:vecA}
A = \Phi dt \;, \quad \Phi = \frac{g_1^2}{4 \pi} \frac{q_{_{\rm BH}}}{r} \;.
\end{equation}

\par The Lagrangian of Eq. (\ref{eq:action}) admits the static and spherically symmetric black hole solution, %\cite{Kramer2003_ExactEFEsols}, 
\begin{equation} \label{eq:metric}
ds^2 = -f(r) dt^2 \; + \; f(r)^{-1} dr^2 \; + \; r^2 \left(d\theta^2 +  \sin^2 \theta  d \phi^2 \right) \;, 
\end{equation}
\noindent written in terms of the Schwarzschild coordinates $(t,r,\theta,\phi)$, with $t \in (-\infty,+\infty)$, $r \in (0,+\infty)$, $\theta \in (0,\pi),$ and $\phi \in (0,2\pi)$. The metric function is defined as
\begin{equation} \label{eq:f}
f(r) = 1 - \frac{2M}{r} + \frac{Q^2}{r^2} - \frac{r^2}{L^2_{dS}}  \;,
\end{equation}
\noindent where the space-time parameters are expressed in terms of length scales \cite{Bekenstein2003_BHinfo} $M$, $Q$, and $L_{dS}$ \cite{AntoniadisBenakli2020_WGCdS}:
\begin{eqnarray}
Gm_{_{\rm BH}} = \frac{\kappa^2}{8 \pi}m_{_{\rm BH}} &=& M \;, \\
\frac{G}{4 \pi }g^2_1 q_{_{\rm BH}} = \frac{\kappa^2}{32 \pi^2 }g^2_1 q_{_{\rm BH}} &=&  Q^2 \;, \\
\sqrt{\frac{3}{\Lambda}} &=& L_{dS} \;.
\end{eqnarray}

\par From the four real roots of Eq. (\ref{eq:f}), we can identify three Killing horizons: the Cauchy horizon $r_-$, the event horizon $r_+$, and the cosmological horizon $r_c$, where
\begin{equation} \label{eq:RNdShorizons}
0 < r_- \leq r_+ \leq r_c \leq L_{dS} < \infty \;.
\end{equation}
\noindent The fourth (and unphysical) root is given by $r_0 = -(r_- + r_+ + r_c)$. These definitions allow for an alternate expression of the metric function, 
\begin{equation} \label{eq:fhor}
f(r) = \frac{1}{r^2 L^2_{dS}}(r-r_-)(r-r_+)(r_c-r)(r-r_0) \;.
\end{equation}

\par For asymptotically flat space-time, the RN black hole develops a degenerate horizon when $M = Q$. This is not the case when $\Lambda >0$, as we shall illustrate explicitly in Fig. \ref{fig:sharkfin}, where there exist valid RNdS black hole solutions for which $Q > M$.

\subsubsection{RNdS black hole phase space: \label{subsec:RNdSphasespace}}
\par In order to delineate the RNdS $(M,Q)$ phase space, we begin with the polynomial
\begin{equation}
\Pi(r) \equiv -r^2 f(r) = -r^2 +2Mr -Q^2 + L^2_{dS} r^4 \;.
\end{equation}
\noindent We then determine the discriminant of this polynomial,
\begin{equation} \label{eq:Det}
\Delta \equiv -16 L^{-2}_{dS} \left[27 M^4 L^{-2}_{dS} - M^2(1 + 36Q^2L^{-2}_{dS}) + (Q + 4Q^3L^{-2}_{dS})^2 \right] \;.
\end{equation}

\noindent By setting $L^2_{dS}=1$, and plotting $\Delta = 0$ we obtain Fig. \ref{fig:sharkfin}. Note also that if we set $\Delta = 0$ and solve for $M$ in terms of $\Lambda$ and $Q$, we reproduce the known analytical bound on $M^2 \Lambda$ \cite{Romans1991_ColdLukewarmRNdS,
Bousso1996_Nariai},
\begin{equation} \label{eq:Mlimit}
M^2 \Lambda \leq \frac{1}{18} \left[ 1 + 12 Q^2 \Lambda + (1 - 4 Q^2 \Lambda)^{3/2}  \right] \;.
\end{equation}
\noindent In Fig. \ref{fig:sharkfin}, where we parametrise $L^{-2}_{dS} = \Lambda/3 =1$, the tip of the sharkfin (Point $U$) corresponds to the most massive black hole solution; there, $M=\sqrt{2/27}$. Traditionally \cite{Romans1991_ColdLukewarmRNdS,Bousso1996_Nariai}, the constraint on the black hole charge in four dimensions comes from the Bogomoln'yi bound for small values of $M^2 \Lambda$ \cite{Romans1991_ColdLukewarmRNdS},
\begin{equation} \label{eq:Qlimit}
\frac{Q^2}{M^2} \lesssim 1 + \frac{1}{3} (M^2 \Lambda) + \frac{4}{9} (M^2 \Lambda)^2 + \frac{8}{9} (M^2 \Lambda)^3 + \mathcal{O}(M^8 \Lambda^4) \;.
\end{equation} 
\noindent For $\Lambda = 3$ and $M=\sqrt{2/27}$, Eq. (\ref{eq:Qlimit}) approximately gives the result $Q = 1/\sqrt{12}$, which is obtained by solving for $Q$ from $\Delta = 0$ and is observed at the tip of the sharkfin. The presence of the cosmological constant raises the upper-bound of the black hole charge-mass ratio from $Q/M \leq 1$ in the asymptotically-flat case to $Q/M \leq 1.06067$ in the asymptotically-de Sitter case.

\par Finally, observe that if we set $L^2_{dS}=1$ and $\underline{\Delta} = -\Delta /16$, we can write the complicated analytical expressions corresponding to the black hole horizons as functions of $M$ and $Q$:
{\setlength{\mathindent}{1cm}
\begin{equation}
\underline{r}_-  =  - a + b \;, \quad
\underline{r}_+  =  + a - b \;, \quad
\underline{r}_c  =  + a + b \;, \quad
\underline{r_0}  =  - a - b \;. \label{eq:radii}
\end{equation}  
}
In this convention, underlined quantities are derived from the parametrised Eq. (\ref{eq:Det}). Here,
{\setlength{\mathindent}{1cm}
\begin{equation}
a=\frac{1}{2 \sqrt{3}} \sqrt{\frac{(1 + X)^2-12 Q^2}{X}}\,  \quad , \quad
b=\frac{1}{2} \sqrt{\frac{4}{3} - \frac{1 - 12 Q^2}{3X} - \frac{X}{3} + \frac{2M}{a}} \;,
\end{equation}
}
and 
\begin{equation}
X = \left(-1 + 54M^2 -36Q^2 - 2\sqrt{27} \sqrt{ \underline{\Delta}} \right)^{1/3} \;.
\end{equation} 

\par Let us now consider the phase space itself. The boundary of Fig. \ref{fig:sharkfin} corresponds to extremised
 conditions\footnote{For clarity, we use the term \enquote{extremise} to refer to black holes with degenerate horizons and \enquote{extremal} to refer to the specific case where inner and outer horizons coincide, $r_-=r_+$.}. At Point $O$ lies pure de Sitter space $(Q=M=0)$. Along the Line $ON$ there is the Schwarzschild de Sitter family of solutions $(Q=0, M>0),$ with the uncharged Nariai case at Point $N$. The Nariai limit, discussed in \ref{app:Nariai}, corresponds to the upper limit of the black hole mass in de Sitter space-time. The charged Nariai branch, for which $r_+ \sim r_c$, extends along Line $NU$, and represents the upper mass limit of the charged black hole. The opposite branch of Line $OU$ corresponds to black hole solutions for which $r_- = r_+$. On this branch, $Q \sim Q_{ext} \sim r_+$, with $Q_{ext}$ defined in Ref. \cite{DiasReallSantos2018_SCC} as 
\begin{equation} \label{eq:qext}
Q_{ext} \equiv y_+ r_c \sqrt{\frac{1 + 2y_+}{1+2y_+ + 3y_+^2}} \;, \quad y_+ = \frac{r_+}{r_c} \;.
\end{equation}
\noindent These branches terminate in Point $U$, where $r_- = r_+ = r_c$ and the local geometry is $\mathbb{M}_2 \times \mathbb{S}^2$.

\subsubsection{Black hole thermodynamics of the RNdS black hole phase space: \label{subsec:RNdSphasespace2}}

\par For each non-degenerate horizon specified in Eq. (\ref{eq:RNdShorizons}), we can calculate a surface gravity $\kappa_i$ and an associated Hawking temperature $T_i$ \cite{Hawking1974_HawkT}. For $i \in \{-,+,c\}$, we define these as
\begin{equation} \label{eq:kappa}
\kappa_i = \frac{1}{2}  \frac{d}{dr} f(r) \bigg \vert_{r=r_i} \;, \quad \quad T_i = \frac{\kappa_i}{2 \pi} \;.
\end{equation}
\noindent Explicitly, the surface gravities corresponding to each horizon are

\begin{eqnarray}
\kappa_- &=&  - \frac{(r_+ - r_-)(r_c - r_-)(r_- - r_0)}{2r_-^2 L^2_{dS}} \label{eq:km} \;, \\
\kappa_+ &=&  + \frac{(r_+ - r_-)(r_c - r_+)(r_+ - r_0)}{2r_+^2 L^2_{dS}} \label{eq:kp}  \;, \\
\kappa_c &=&  -\frac{(r_c - r_-)(r_c - r_+)(r_c - r_0)}{2r_c^2 L^2_{dS}} \label{eq:kc}  \;,
\end{eqnarray}
\noindent with $|\kappa_-| > |\kappa_+|$ \cite{MossMyers1998_CC}. For non-extremal black holes, the Hawking radiation emanates from each horizon at a different temperature. However, degenerate horizons have vanishing surface gravities. Since $T_i=0$ in these cases, these black holes are classified as \enquote{cold}. For asymptotically flat space-time, the RN black hole develops a degenerate horizon when $M=Q$ and $\kappa_- = \kappa_+ = 0$. For the RNdS case, $\kappa_- = \kappa_+ = 0$ when $Q_{ext} \sim r_+$ for $Q_{ext}$ defined in Eq. (\ref{eq:qext}), as suggested in Ref. \cite{DiasReallSantos2018_SCC}.

\begin{figure}[t!]
\centering
\includegraphics[width=0.6\linewidth]{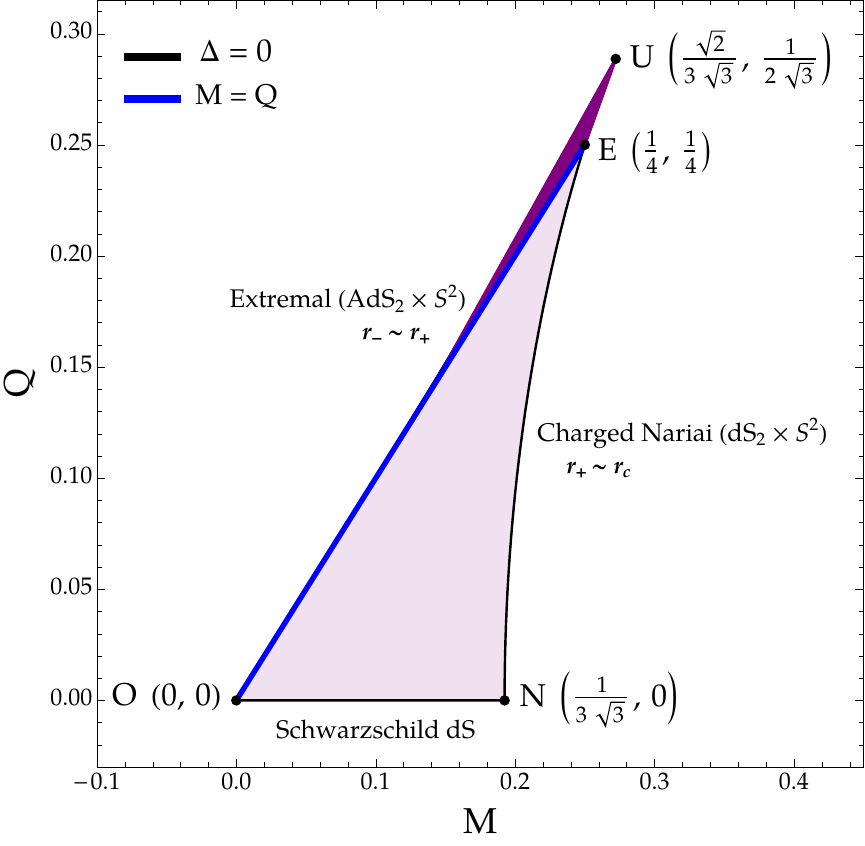}
\caption{ \textit{A two-dimensional projection of the four-dimensional RNdS black hole parameter space for $L^{-2}_{dS} = \Lambda/3 =1$. Dark (light) shading corresponds to cold (warm) black holes. Diagram originally sketched in Ref. \cite{Chrysostomou:2024inc}, inspired by Figure 1 of Ref. \cite{vanRiet2019_FLevapBHdS}.
}}
\label{fig:sharkfin}
\end{figure}

\par Thermal radiation isotropically pervades de Sitter space-time \cite{GibbonsHawking1977_BHTherm}. As such, only two black hole solutions represent thermal equilibrium in the RNdS case: the \enquote{lukewarm} solutions for which $M=Q$, and the \enquote{charged Nariai} families corresponding to the condition $r_+ = r_c$ \cite{Romans1991_ColdLukewarmRNdS}. Black holes undergoing charge emission slowly evolve towards these lukewarm solutions \cite{Romans1991_ColdLukewarmRNdS,
Brill1993_RNdSextrema,
Mann1995_ChargedBHpairs,
Bousso1999_QuantumStructredS}. With this in mind, we can divide the phase diagram of Fig. \ref{fig:sharkfin} into two regions: the \enquote{colder} $OEU$ region, where $Q>M$, and the warmer $OEN$ region, where $Q<M$.

\par Black hole solutions in the shaded $Q > M$ region are colder than the lukewarm solutions and absorb radiation from the cosmological horizon until they become lukewarm. This is evidenced also by $T_c <0$. The $Q > M$ region is bounded by the $OU$ line, where if $Q$ increases with respect to $M$, the inner and outer black hole horizons eventually coincide, leading to the extremal cold RNdS black hole condition.

\par Below the lukewarm line is the $Q< M$ region. In this region, black holes are warmer than the cosmological horizon, and will evaporate until they become lukewarm. Note that because these black holes are stabilised by the Gibbons-Hawking radiation from the cosmological horizon and cannot become arbitrarily cold. The region is bounded by the $NU$ line, where $r_+ \sim r_c$. The outer and cosmological horizons are in thermal equilibrium, leading to the charged Nariai solution, which has a black hole mass that is maximal for a given charge \cite{Bousso1996_Nariai}.

\par Finally, let us close this section with a few additional constraints that can be derived for the RNdS black hole space-time. As explained in Appendix A of Refs. \cite{refNatarioSchiappa,refIKrn}, we can examine the extremised conditions with respect to the black hole charge-mass relationship, and the constraint on $L_{dS}$,
\begin{equation}
    0 < L^2_{dS_{\pm}} =  \frac{1}{2}  \frac{(3M \pm \sqrt{9M^2-8Q^2})^3}{M \pm \sqrt{9M^2-8Q^2}} \;.
\end{equation}
Here, we shall observe how this corresponds to the region within the sharkfin diagram. 

\begin{outline}
    \1[$(i)$] For $Q^2 > M^2$, above the $M=Q$ line, we have a number of extremising configurations: 
    \2[$(a)$] We can impose extremal conditions on $Q$, such that $Q^2 = 9M^2/8$. This yields the ``RN-type" extremal condition that extremises the cosmological constant at the upper value, such that $L^2_{dS} = L^2_{dS_{+}}$, i.e. $\Lambda~=~2/9M^2$. This yields the \enquote{ultra-extremal}/\enquote{ultracold} case \cite{Romans1991_ColdLukewarmRNdS,Brill1993_RNdSextrema} at Point $U$, where Hawking temperature goes to zero and the local geometry is $\mathbb{M}_2 \times \mathbb{S}^2$. There is only one real positive root,
\begin{equation} \label{eq:ultracold}
r_- = r_+ = r_c = \frac{3M}{2} \;.
\end{equation}
  \2[$(b)$] Alternatively, we can extremise $\Lambda$. For the $L^2_{dS} =L^2_{dS_-}$ case, we have the ``RN-type'' extremal condition that gives two positive roots; one at $r=r_c$ and a degenerate horizon at
\begin{equation} 
r_- = r_+ = \frac{3M - \sqrt{9M^2-8Q^2}}{2}  \;. 
\end{equation}
This is the ``cold” black-hole case \cite{Romans1991_ColdLukewarmRNdS,Brill1993_RNdSextrema} on Line $OU$.
\1[$(ii)$] For $Q^2 \leq M^2$, under the $M=Q$ line, the ``dS-type extremal condition" $L^2_{dS} = L^2_{dS_{+}}$ gives two positive roots,  
\begin{equation} 
r_+ = r_c = \frac{3M + \sqrt{9M^2-8Q^2}}{2}  \;. \end{equation}
This is the ``extreme" or ``marginal" naked-singularity case \cite{Romans1991_ColdLukewarmRNdS,Brill1993_RNdSextrema}, falling on Line $NU$. 
\end{outline}

\par From this overview of the RNdS phase space, we have reviewed a number of limits on the mass and charge of the black hole with respect to a de Sitter radius of $L_{dS}=1$. The phase space diagram of Fig. \ref{fig:sharkfin} shall serve as the foundation of our QNM analyses of Section \ref{chap:RNdSQNMs}. 

%%%%%
%SECTION III
%%%%%

\renewcommand{\theequation}{\arabic{section}.\arabic{subsection}.\arabic{equation}}
\section{Probing the RNdS black hole space-time with QNMs \label{chap:RNdSQNMs}}  

\subsection{Setup for charged massive scalar QNMs in the RNdS space-time \label{sec:QNMsRNdS}}
\par To begin our discussion on the QNMs of a charged and massive scalar test field within the RNdS black hole space-time, we consider the generic action for the Einstein-Hilbert-Maxwell system of Eq. (\ref{eq:action}) minimally coupled to a complex scalar test field charged under $U(1)$:

\begin{equation}
	S =  \int d^4x \sqrt{-g} \left[ \frac{1}{2\kappa^2}\left(R - 2\Lambda  \right) - \frac{1}{4g_1^2} F_{\mu \nu} F^{\mu \nu} \right] +  \int d^4x \sqrt{-g} \mathcal{L}_{sc.} \;.
\end{equation}
\noindent As before, we take $U(1)$ to be electromagnetism. The Lagrangian $\mathcal{L}_{sc.}$ represents the complex scalar field, which is itself a linear combination of two real scalar fields $\Phi_1$ and $\Phi_2$, such that $\Phi = \Phi_1 + i\Phi_2$. The full Lagrangian \cite{HawkingEllis1973_LSSUbook,
BritoCardosoPani2020_Superradiance} is then given by
\begin{equation}
\mathcal{L}_{sc.} = - \frac{1}{2} \left( \mathcal{D}_{\mu} \Phi \right)^{\dagger} \left( \mathcal{D}^{\mu} \Phi \right) - \frac{1}{2}\mu^2 \Phi^{\dagger} \Phi 
\;. 
\end{equation}
Here, $\mathcal{D}_{\mu} = (\partial_{\mu} - iqA_{\mu})$. The interaction of the charged scalar field with the external electromagnetic field of the black hole, $F_{\mu \nu} = \partial_{\mu} A_{\nu} - \partial_{\nu} A_{\mu}$, is introduced through the minimal coupling prescription. Under this prescription $\partial_{\mu}$ is replaced by its covariant counterpart, $\mathcal{D}_{\mu}$. Recall that in this stationary black hole context, the only nonzero component of $A_{\mu}$ is $A_t(r)~=~-~Q/r~dt$, the electrostatic four-potential of the black hole.

\par The full, non-linear evolution of the system can be described using the corresponding equations of motion. These separate into the massive, charged Klein-Gordon equation in curved space-time and the Einstein field equations:
\begin{equation}
    \nabla_{\mu} \nabla^{\mu} \Phi - \mu^2 \Phi = 0 \;, \quad G_{\mu \nu} + \Lambda g_{\mu \nu} = 8 \pi G T_{\mu \nu} \;,
\end{equation} 
\noindent for which the stress-energy tensor becomes quadratic in $\Phi$ for this particular model. Higher-order perturbations in the scalar field induce changes in the space-time geometry, as well as in the vector potential \cite{BritoCardosoPani2020_Superradiance}. Following the usual conventions \cite{refBertiCardoso}, we avoid these complications by introducing the linear approximations for the fields $\Phi$ and $g_{\mu \nu}$,
\begin{equation} \label{eq:perts}
    g'_{\mu \nu} = g^{\rm BH}_{\mu \nu} + \delta_{\mu \nu} \;, \quad \Phi'=\Phi^{\rm BG} + \Psi \;.
\end{equation}
\noindent The unperturbed fields, $g^{\rm BH}_{\mu \nu}$ and $\Phi^{\rm BG}$, are referred to as the \enquote{backgrounds}. The \enquote{perturbations}, $\delta_{\mu \nu}$ and $\Psi$, are considered to be very small. If we substitute $g'_{\mu \nu}$ and $\Phi'$ (with $\Phi^{\rm BG}=0$) and linearise the system of equations with respect to $\delta_{\mu \nu}$ and $\Psi$, we find that $\delta_{\mu \nu}$ and $\Psi$ decouple. The metric fluctuations for $\delta_{\mu \nu}$ can then be set to zero and $g^{\rm BH}_{\mu \nu}$ satisfies the vacuum Einstein field equations. In this way, the gravitational sector can be described by the vacuum solution $R_{\mu \nu}=0$, and the backreaction emergent at quadratic order in $\Phi$ is neglected.

\par Let us proceed to the equation of motion for $\Psi$, 
\begin{equation} \label{eq:KGrnds}
\frac{1}{\sqrt{-g}} \left( \partial_{\mu} - iq A_{\mu} \right)\left( \sqrt{-g} g^{\mu \nu} \left( \partial_{\nu} - iq A_{\nu} \right) \Psi \right) = \mu^2 \Psi \;.
\end{equation}
\noindent Recall that we can formulate an ansatz for $\Psi$ derived from the symmetries of the fixed background space-time. In the RNdS case (static, non-rotating, and spherically-symmetric), the wave-function is written in variable-separable form,
\begin{eqnarray} \label{eq:ansatzrnds}
    \Psi_{n \ell } (t,r,\theta,\phi) = \sum_{n=0}^{\infty}\sum_{\ell=0}^{\infty} \frac{\psi_{n\ell } (r)}{r} \; Y_{\ell } (\theta,\phi) \; e^{-i \omega_{n \ell } t} \;.
\end{eqnarray}
As explained in the introduction, we are concerned only with the $n=0$ \enquote{fundamental mode}, representing the least-damped and thus longest-lived QNM. The angular contribution is expressed using spherical harmonics, for which  $\ell$ represents the angular momentum number. The spherical harmonic function $Y_{\ell } (\theta, \phi )$ satisfies 
\begin{equation}
\nabla^2 Y_{\ell } (\theta, \phi ) = -\frac{\ell (\ell + 1)}{r^2} Y_{\ell } (\theta, \phi) \;.
\end{equation}

\noindent Since the black hole is static, the corresponding ordinary differential equations are time independent. Consequently, the defining QNM behaviour is then fully encapsulated by the radial component. For convenience, we drop the subscripts and write
{\setlength{\mathindent}{1cm}
\begin{equation}
\frac{d}{dr}\left(r^2 f(r)\frac{d \psi}{dr}\right)+\left(\frac{r^2(\omega+qA_t(r))^2}{f(r)}-\ell(\ell+1)-\mu^{2}r^2 \right) \psi(r)=0\,. \label{eq:radial}
\end{equation}%
}
\noindent Then, redefining $\psi(r)$ as $\psi(r)=\varphi(r)/r$ and employing the tortoise coordinate $r_{\star}$, we obtain
{\setlength{\mathindent}{1cm}
\begin{equation} \label{eq:odeRNdS}
\frac{d^{2}\varphi(r_{\star})}{dr_{\star}^{2}} + \left[ \omega^2 -V(r) \right] \varphi(r_{\star})=0 \;,
 \end{equation}
\begin{equation}\label{eq:potRNdS}
\mathrm{with} \;\;\; \; V(r)  = f(r) \left[ \frac{\ell (\ell + 1)}{r^2}  +   \frac{f^\prime(r)}{r} + \mu^2 \right] -2\omega q A_t(r)-q^2A_t(r)^2 \;. 
\end{equation}
}
\noindent Note that in the RNdS space-time,
where $r_{\star}=r_{\star}(r)$ serves as a bijection from $(r_+,r_c)$ to $(-\infty,+\infty)$, the general tortoise coordinate becomes
\begin{equation}
r_{\star} (r)=\int \frac{dr}{f(r)} = \sum^{4}_{i=1} \frac{1}{2\kappa_i} \ln \left(1- \frac{r}{r_i} \right) \, ,
\end{equation}
\noindent if we set $r_{\star}(r=0)=0$. Here, $\kappa_i$ is the surface gravity as defined in Eq. (\ref{eq:kappa}).
\par For a massive charged scalar field in a RNdS background, the QNM problem to solve can be written in the form of a wave equation, 
{\setlength{\mathindent}{1cm}
\begin{equation} \label{eq:odefullRNdS}
\frac{d^{2}\varphi(r_{\star})}{dr_{\star}^{2}} + \left[ \left( \omega - \frac{qQ}{r} \right)^2 - f(r) \left[ \frac{\ell (\ell + 1)}{r^2} + \frac{f^{\prime}(r)}{r} + \mu^2 \right]  \right] \varphi(r_{\star})=0 \;,
 \end{equation}
 }
\noindent with a squared mass term $\mu^2$ and where $\omega$ is shifted by the $qQ/r$ term. The QNM boundary conditions undergo a similar shift,
{\setlength{\mathindent}{1cm}
\begin{equation} \label{eq:BCdS}
\varphi(r_{\star}) \sim 
\cases{e^{-i \left(\omega - \frac{qQ}{r_+} \right)r_{\star}} \;, \quad \quad r \rightarrow r_+ \;\; (r_{\star} \rightarrow - \infty) \;,\\
e^{+i\left(\omega - \frac{qQ}{r_c} \right)r_{\star}} \;, \quad  \quad r \rightarrow r_c \; \;\; (r_{\star} \rightarrow + \infty) \;,}{}
\end{equation} 
}
\noindent with radiation purely outgoing at the de Sitter horizon  \cite{KonoplyaZhidenko2014_RNdSrprc,Hod2018_RNdS,
DiasReallSantos2018_SCC,
DiasSantos2020_RNdSinstability}.

 \par These QNM boundary conditions are applied to perturbations exterior to the event horizon. Recall that, classically, the event horizon serves as a unidirectional causal boundary: energy may cross the boundary to enter into the black hole interior but cannot escape the black hole thereafter. Furthermore, energy cannot enter into the system from beyond the de Sitter horizon. As such, the system is not time-symmetric; the QNM eigenvalue problem at hand is non-Hermitian and the eigenvalues (i.e. the QNFs) are complex. The real part of the complex QNF $\omega$ represents the physical oscillation frequency and the imaginary part is related to the inverse damping rate of the oscillations. The corresponding eigenfunctions $\varphi$ are not normalisable and do not form a complete set (see reviews \cite{refNollert1999,refBertiCardoso,refKonoplyaZhidenkoReview}).

\par A host of QNM techniques have been developed to confront these technical difficulties, several of which are semi-classical methods, informed by the Schutz-Iyer-Will WKB-based technique \cite{refBHWKB0,refBHWKB0.5,refBHWKB1}, that take advantage of the QNM problem's likeness to a scattering problem. Specifically, these require a bell-shaped confining potential and particular asymptotic behaviour (i.e. $V(r_{\star})$ tends towards constant values as $r_{\star} \rightarrow \pm \infty$) for their applicability \cite{Konoplya2019_recipes}. The scalar field parameters strongly influence the shape of the potential and, in turn, the nature of the QNFs. As described in the introduction, the damping of the QNFs decreases with increasing field mass. In this way, for particular values of $\mu$ and $\ell$, QNFs within the Schwarzschild black hole space-time enter the \enquote{quasiresonance} regime and become arbitrarily long-lived.

\par As we explored in Ref. \cite{Chrysostomou2023_EPJC}, an analysis of the behaviour of the potential allows us to estimate the values of $\mu$ and $\ell$ at which the potential barrier is smoothed out, serving as a rough precursor to the onset of the quasiresonant regime. For scalar perturbations in the RNdS case, there exists a known instability for small scalar field charge and vanishing multipolar number \cite{ZhuZhang2014_RNdSinstability}, with a superradiant origin \cite{KonoplyaZhidenko2014_RNdSrprc}. For this reason, the $\ell = 0$ and $\ell >0$ cases must be treated independently: we include here a brief review of the former but focus our attention on the latter.

We shall outline the necessary QNF condition for superradiant instabilities and demonstrate how the shape of the potential suggests superradiant amplification of incident waves reflected off of the potential in Section \ref{subsec:super}. Thereafter, we shall describe the behaviour of the potential for $\ell >0$ within the \enquote{sharkfin}, which informs our subsequent QNF analyses in Section \ref{sec:QNFspecRNdS}.

\subsubsection{Superradiance  for $\ell=0$ in the RNdS black hole space-time: \label{subsec:super}}

\begin{figure}[t]
    \centering
        \includegraphics[width=0.5\linewidth]{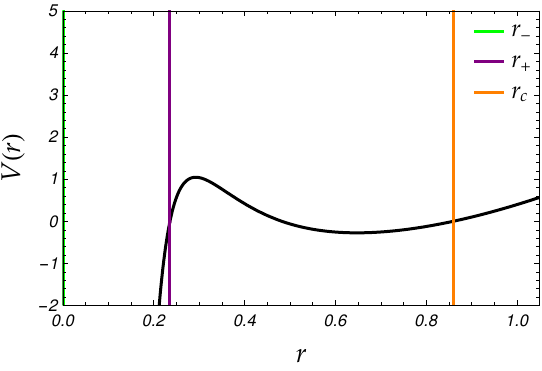}
     \caption{\textit{For $L^2_{dS} =1$, $M=0.112$, and $Q=0.016$, we plot the scalar QNM potential with $\mu=q=0.1$ and $\ell=0$. Observe the \enquote{valley} following the barrier potential, indicative of superradiant amplification.}} \label{fig:potential-zeroL} 
\end{figure}

\par An exponentially growing mode can be caused by superradiance. As mentioned earlier, when $\mathfrak{I}m \{ \omega \} \approx 0$, QNMs enter the \enquote{quasiresonance regime} in which modes become arbitrarily long-lived. This occurs when the field mass reaches some critical value, such that the damping rate asymptotically approaches zero and the QNM behaviour resembles that of a bound state \cite{Ohashi2004_MassiveScalarQ}. The mass responsible for this decay in the damping rate is associated with a local minimum generated far from the black hole. Waves scattered off of the black hole potential barrier become reflected and amplified within this \enquote{valley} \cite{Furuhashi:2004jk}. This was observed in the case of the RNdS black hole for the $\ell = 0$ spin-0 QNM for a vanishing $\mu$ in Ref. \cite{ZhuZhang2014_RNdSinstability}. A number of studies scrutinising the parameter space of unstable QNFs within a RNdS background followed e.g. Refs. \cite{KonoplyaZhidenko2013_dRNdSinstability,
KonoplyaZhidenko2014_RNdSrprc,
DiasEperonReallSantos2018_SCC}, confirming instability for a vanishing field mass, as well as a small field charge and a small cosmological constant. This is illustrated in Fig. \ref{fig:potential-zeroL}.

\par Let us consider a black hole scattering scenario. In the RNdS case, the ingoing wave from $r=r_c$ partially passes through the potential barrier, passing $r = r_+$ to fall inside the event horizon, while the rest is reflected from the potential barrier back towards the cosmological horizon $r= r_c$. The RNdS boundary conditions of Eq. (\ref{eq:BCdS}) are modified to those of a scattering problem, such that
{\setlength{\mathindent}{1cm}
\begin{equation} \label{eq:BCdSrt}
\varphi \sim 
\cases{\mathcal{T} e^{-i \left(\omega - \frac{qQ}{r_+} \right)r_{\star}} \;, \hspace{2.6cm} \quad r \rightarrow r_+ \;\; (r_{\star} \rightarrow - \infty) \;,\\
e^{-i \left(\omega - \frac{qQ}{r_c} \right)r_{\star}} + \mathcal{R} e^{+i\left(\omega - \frac{qQ}{r_c} \right)r_{\star}} \;, \quad \quad r \rightarrow r_c \;\; (r_{\star} \rightarrow + \infty) \;.}{}
\end{equation}
}

\noindent Here, $\mathcal{R}$ is the amplitude of the reflected wave (i.e. the reflection coefficient), and $\mathcal{T}$ is the transmitted-wave amplitude (i.e. transmission coefficient). We have set the incident amplitude $\mathcal{I}$ to unity. Superradiance corresponds to $\mathcal{R}>1$, i.e. the amplitude of the reflected wave exceeding that of the incident wave. For the linear-independent solutions, the Wronskian will be constant, and the relationship between reflection and transmission coefficients becomes 
\begin{equation}
1-|\mathcal{R}|^2 = \frac{\omega - qQ/r_+}{\omega - qQ/r_c} |\mathcal{T}|^2 \;.
\end{equation} 
\noindent With this expression in mind, a necessary condition for superradiance can be derived from Eq. (\ref{eq:odefullRNdS}). Following Ref. \cite{KonoplyaZhidenko2014_RNdSrprc}, this is given by
\begin{equation} \label{eq:superradiance}
\frac{qQ}{r_c} < \mathfrak{R}e \{ \omega \} < \frac{qQ}{r_+}
\;.
\end{equation}
\noindent In Ref. \cite{KonoplyaZhidenko2014_RNdSrprc}, it was observed that the growing modes satisfied Eq. (\ref{eq:superradiance}). However, some stable modes also satisfied this superradiance condition. The authors there concluded that superradiance can imply instability but does not necessarily prove instability. In other words, superradiant modes are not necessarily unstable modes. 

\par A superradiant amplification of reflected charged perturbations within RNdS space-times can be inferred directly from a study of the QNM effective potential. In Fig. \ref{fig:potential-zeroL}, we plot the $\ell = 0$ case for $L^2_{dS} =1$, $M=0.112$, $Q=0.016$, and $\mu=q=0.1$. These parameters correspond to a central position in the phase space of Fig. \ref{fig:sharkfin}. There is a single \enquote{peak} in the barrier potential, the local maximum, followed by a local minimum. A wave scattering off the potential barrier will become amplified in this \enquote{valley}, resulting in a superradiance that destabilises the black hole. For fixed $Q$, increasing $M$ shrinks the amplitude of the potential.

\par {\color{black}{As we shall see in the next section, increasing $\ell$ or $\mu$ lifts the potential, restoring stability. We note with interest that for the extremised conditions showcased in Figs. \ref{fig:RNdSpotential-chNariai} and \ref{fig:RNdSpotential-cold}, where the local maximum is suppressed and a local minimum follows, the \enquote{valley} appears beyond the physically-relevant domain $r_+ < r < r_c.$ As confirmed in Ref. \cite{Konoplya2024_TwoRegimes}, a single peak remains in the region defined as the black hole exterior. 

\par Finally, we note that for a harmonic time dependence $\psi \sim e^{-i \omega t}$, a negative imaginary component reveals an exponentially decaying system, in accordance with a return to an equilibrium state \cite{Vishveshwara1970_stability}. An unstable QNM is indicated by $\mathfrak{I}m \{ \omega \} >0$, corresponding to an exponential growth in the oscillations. As we shall discuss in Section \ref{sec:QNFspecRNdS}, the RNdS QNF spectrum is complicated by the introduction of a non-zero $q$ and the consequent breaking of the $\omega \rightarrow -\omega^+$ QNM reflection symmetry. To capture the unstable and/or superradiant behaviour of the imaginary component, we require a more precise method than the WKB-based technique we use here (e.g. direct integration or pseudospectral techniques), as results from WKB-based methods in this case would be misleading. Within the literature, however, there has been no evidence of instability in the RNdS space-time for $\ell \neq 0$. }}

\subsubsection{The behaviour of the potential within the RNdS phase space {\color{black}{for $\ell \geq 1$}}: \label{sec:potRNdS}}

\par Let us consider the QNM potential of a massive and charged scalar test field with respect to key benchmark points within the RNdS phase space of Fig. \ref{fig:sharkfin}. In these sketches of the potential, we demarcate the locus of each horizon as defined by the roots of the metric function for each given benchmark: green, purple, and orange vertical lines for the Cauchy $(r=r_-)$, event $(r=r_+)$, and cosmological $(r=r_c)$ horizons, respectively. As we shall see, a single peak in the potential is defined in each case on the black hole exterior $r \in (r_+ , r_c)$, framed by the purple and orange delineators. The barrier potential established in this context suggests that QNM propagation is maintained and supports the application of WKB-based techniques. {\color{black}{This behaviour of the potential was recently confirmed in Ref. \cite{Konoplya2024_TwoRegimes}, which determined that for general spherically-symmetric black holes, the WKB method is indeed quite accurate for $\Lambda >0$, particularly if $\mu M \gg 1$. Note that this is in direct contrast with the Schwarzschild case studied in Ref. \cite{Chrysostomou2023_EPJC}, where the local maximum of the potential is suppressed by large values of $\mu$ and the WKB method fails to produce reliable results. Once we have discussed the behaviour of the potential, we shall return to this point at the end of this section.}}

\begin{figure}[t]
    \centering
        \includegraphics[width=0.5\linewidth]{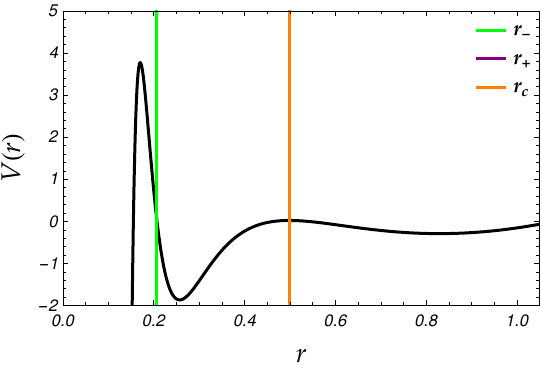}
        \caption{\textit{
        For $M=Q=0.25$  and $L^2_{dS} =1$ (Point $E$ of Fig. \ref{fig:sharkfin}), we plot the scalar QNM potential with $\ell=1$ and $\mu=q=0.1$.}}
        \label{fig:RNdSpotential-chNariai}
\end{figure}

\begin{figure}[t]
    \centering
        \includegraphics[width=0.5\linewidth]{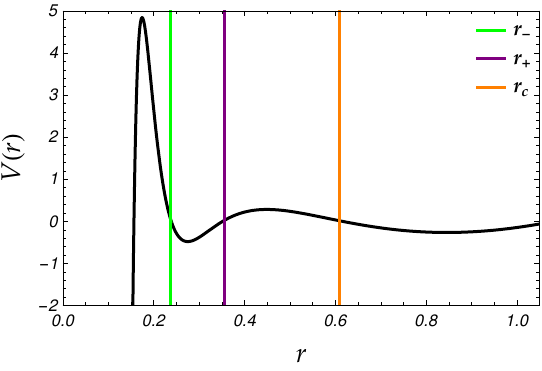}
    \caption{\textit{
    For $M=0.2425$, $Q=0.249$, and $L^2_{dS} =1$ (above the $M=Q$ line of Fig. \ref{fig:sharkfin}), we plot the scalar QNM potential with $\ell=1$ and $\mu=q=0.1$. 
    }}
           \label{fig:RNdSpotential-cold}
\end{figure}

\par In Fig. \ref{fig:RNdSpotential-chNariai}, we plot the potential for $\mu=q=0.1$ and $\ell=1$ for $M=Q=1/4$. This is Point $E$ of Fig. \ref{fig:sharkfin}, corresponding to a charged Nariai case at the maximum point of the $M=Q$ \enquote{lukewarm} line. In setting $r_+ \rightarrow r_c$ (see \ref{app:Nariai} for details on the Nariai limit), there is no physical exterior black hole region; the de Sitter space-time is dominated entirely by the black hole. This renders QNM analyses impossible, as we cannot impose the boundary conditions necessary to isolate the discrete set of frequencies. In this configuration, we see explicitly the effect of the cosmological constant on the parameter space: the $M=Q$ solution, which in asymptotically flat space-times corresponds to a coalescence of the Cauchy and event horizons, is associated with a meeting of the event and de Sitter horizons. Similarly, for the \enquote{ultracold} case (Point $U$ on Fig. \ref{fig:sharkfin}), all horizons converge on this line at $r =3M/2 \sim 0.408$.

\begin{figure}[t]
\centering
    \begin{subfigure}[t]{0.5\textwidth}
        \centering
        \includegraphics[width=0.99\textwidth]{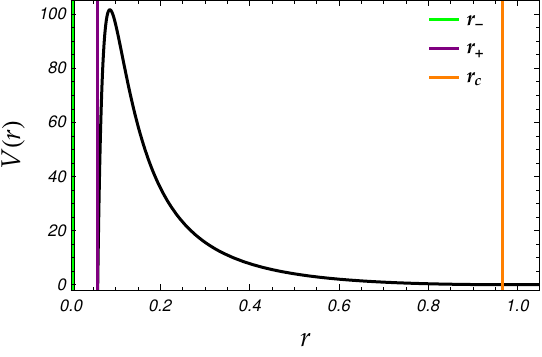}
        \caption{\textit{$M=0.032$ and $Q=0.016$, $V(r_{p})\sim 100$.}}
    \end{subfigure}%
    ~ 
\begin{subfigure}[t]{0.5\textwidth}
        \centering
        \includegraphics[width=0.99\textwidth]{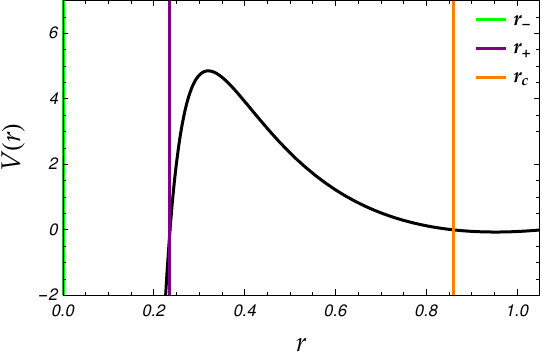}
        \caption{\textit{$M=0.112$ and $Q=0.016$, $V(r_{p})\sim 5$.}}
    \end{subfigure}\\
    \vskip 0.3cm
    ~ 
    \begin{subfigure}[t]{0.5\textwidth}
        \centering
        \includegraphics[width=0.99\textwidth]{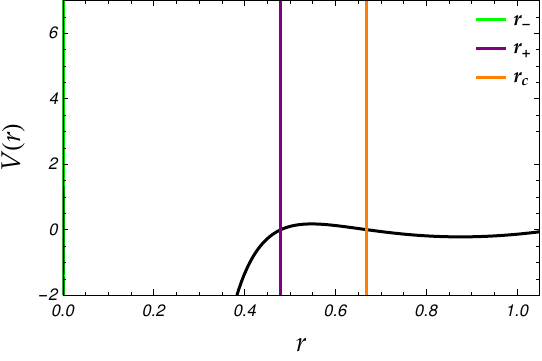}
        \caption{\textit{$M=0.185$ and $Q=0.016$, $V(r_{p}) < 0.5$. \label{fig:subfigc}}}
    \end{subfigure}
    ~ 
 \vskip 0.3cm
    \caption{\textit{The evolution of the QNM scalar potential within Fig. \ref{fig:sharkfin} for $L^2_{dS}=1$, $\ell=1$, and $\mu=q=0.1$. 
    }}
   \label{fig:RNdSpotentials}
\end{figure}

\par To our knowledge, there has been little discussion in the literature on the nature of QNMs within \enquote{cold} black holes, such that $Q>M$. To determine if there is a significant behavioural shift upon crossing the \enquote{lukewarm} line, we plot the potential for $M=0.2425$ and $Q=0.249$ in Fig. \ref{fig:RNdSpotential-cold} for these same field parameters of $\ell =1$ and $\mu=q=0.1$. The barrier potential is present, but suppressed. Interestingly, for the slightly \enquote{warmer} $M>Q$ solution corresponding to $M=0.2425$ and $Q=0.239$, the event and de Sitter horizons are relatively closer and the potential's amplitude becomes more suppressed.

\par
\begin{figure}[t]
\centering
    \begin{subfigure}[t]{0.47\textwidth}
        \centering
        \includegraphics[width=\textwidth]{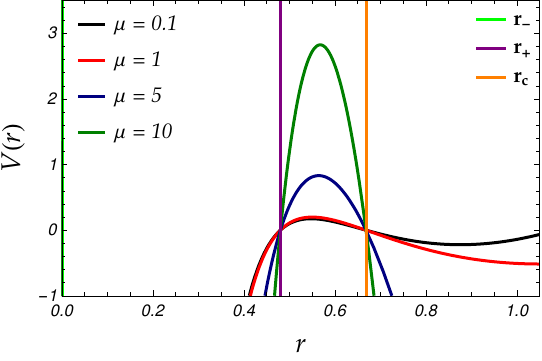}
        \caption{\textit{For $q=0.1$, $V(r_{p})$ scales with $\mu$.
        \label{fig:LargeMassQNM}}}
    \end{subfigure}
    ~ 
        \begin{subfigure}[t]{0.47\textwidth}
        \centering
        \includegraphics[width=\textwidth]{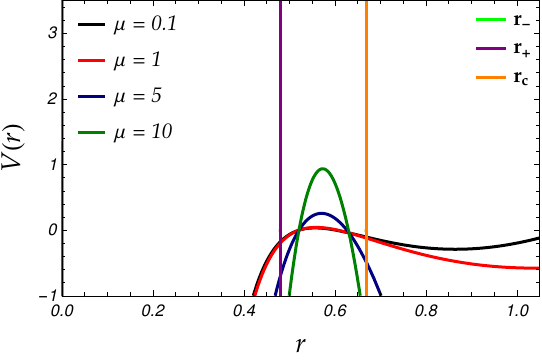}
        \caption{\textit{For $q=5$, the amplitude is suppressed.}}
    \end{subfigure}
    ~ 
 \vskip 0.3cm
    \caption{\textit{{\color{black}{The influence of scalar field parameters on the QNM scalar potential within Fig. \ref{fig:sharkfin} for $L^2_{dS}=1$, $M=0.185$, $Q=0.016$, and $\ell=1$. }}}}
   \label{fig:RNdSpotentials2muq}
\end{figure}

 After exploring the full available phase space depicted in Fig. \ref{fig:sharkfin}, we can summarise the general behaviour of the potential. As a visual aide, we illustrate examples of these trends in Figs. \ref{fig:RNdSpotentials} and \ref{fig:RNdSpotentials2muq}. For a fixed $Q$, the peak of the amplitude of the potential $V(r_{p})$ is largest near Point $O$ and decreases towards $U$, remaining largest along the extremal branch $OU$. The amplitude of the potential decreases significantly as we increase $M$ from the near-extremal $OU$ branch to the near-Nariai $NU$ branch. Despite this, the potential retains its shape on $r_+ < r < r_c$. We can also identify the effect of the scalar field parameters on the barrier potential. For example, raising $\ell$ increases the magnitude of $V(r)$, as expected. {\color{black}{From Fig. \ref{fig:RNdSpotentials2muq}, we see that even for the suppressed peak of Fig. \ref{fig:subfigc}, elevating $\mu$ increases the amplitude of the peak and smoothes out the local minimum. As suggested in Ref. \cite{Konoplya2024_TwoRegimes}, a strong barrier potential emerges in the $\mu M \gg 1$ regime in the region of interest $r_+ < r < r_c$. Raising $q$, on the other hand, suppresses the effective potential.}}

\par {\color{black}{From these observations, the application of QNM techniques dependent on barrier potentials (e.g. WKB-based, potential-based, and photon-orbit techniques) seems permissible throughout the sharkfin (except for the extremised regions), provided the field parameters support a barrier potential.}} Recall from our discussion in Section \ref{subsec:super} that the standard WKB method (e.g. the Schutz-Iyer-Will approach) remains a good approximation provided the local maximum of the potential is \enquote{high enough} for $\ell \geq 1$. Even if a local minimum follows the local maximum, WKB-based methods remain fairly reliable, provided $V(r_{p})$ exceeds the value to which $V(r)$ asymptotes as $r \rightarrow \infty$. In this case, Ref. \cite{Konoplya2019_recipes} claims that neglecting the local minimum does not result in significant error.

\par With these arguments in mind, we shall make use of a modified WKB method in our QNM investigations within the RNdS phase space. This method was established in Ref. \cite{Hatsuda2019_WKB} and employed for scalar QNMs of nonzero mass and charge in the RNdS context in Refs. \cite{Fontana2020_RNanomalous,Papantonopoulos2022}. The value of this method lies in the fact that it allows us to define the position of the potential peak explicitly through Eq. (\ref{eq:rmax}) and to produce the QNF as a series expansion that maintains the space-time and field parameters, $(M,Q,\Lambda)$ and $(\mu,q,\ell)$, respectively, as free parameters. We shall introduce this method in the following section.

\subsection{The QNF spectrum in RNdS space-time \label{sec:QNFspecRNdS}}

\subsubsection{The semi-classical calculation of QNFs in a RNdS background: \label{subsec:WKBbasedTechnique}}
\par In this section, we compute the QNFs using a modified WKB approach applied to the RNdS black hole space-time by Fontana \textit{et al.} \cite{Fontana2020_RNanomalous,Papantonopoulos2022} in the eikonal regime for small $Q/M$ and $qM$. The method is based on a Borel-resummation technique proposed in Ref. \cite{Hatsuda2019_WKB} that exploits the well-established relationship between the bound states of anharmonic oscillators \cite{BenderWu1969_AnharmonicOscillator} and the QNFs of black holes \cite{PoschlTellerMethod,refFerrMashh1,
refFerrMashh2}. 

\par %Like the Dolan-Ottewill multipolar expansion method \cite{refDolanOttewill2009}, 
To begin, the QNF is expressed as a series expansion in inverse powers of $L$,
\begin{equation} \label{eq:omegaseries}
\omega = \sum_{k = -1} \omega_k L^{-k} \;, \quad L = \sqrt{\ell (\ell+1)}\; .
\end{equation} 
\noindent This series expansion is then inserted into
\begin{eqnarray} \label{eq:BorelQNF}
\omega &=&\sqrt{V(r^{max}_{\star})-2 i U} \;, \nonumber \\
U &\equiv & U(V^{(2)},V^{(3)},V^{(4)},V^{(5)},V^{(6)})  \;.
\end{eqnarray}
\noindent Here, $V(r^{max}_{\star})$ refers to the peak of the barrier potential, located at 
\begin{eqnarray} 
 r^{max}_{\star} &\approx & r_0 + r_1 L^{-2} + ...\;, \nonumber \label{eq:rmax}\\ 
 V(r^{max}_{\star}) &\approx &  V_0 + V_1 L^{-2} + ... \;, \label{eq:Vmax}
\end{eqnarray}
where subscripts refer to terms in a series expansion around the peak. The explicit expressions for $U$, $V(r^{max}_{\star})$, and $r^{max}_{\star}$ are presented in \ref{app:num}. 
%\begin{widetext}
\par The numbered superscripts in Eq. (\ref{eq:BorelQNF}) refer to derivatives $V^j$, taken with respect to a generalised tortoise coordinate, such that
{\setlength{\mathindent}{1.5cm}
\begin{equation} \label{eq:Vj}
V^{j} = \frac{d^j V(r^{max}_{\star})}{dr^j} = f(r) \frac{d}{dr} \left[ f(r) \frac{d}{dr} \left[... \left[f(r) \frac{d V(r)}{dr} \right]... \right] \right]_{r \rightarrow r^{max}_{\star}} \;.
\end{equation}
}
%\end{widetext}
\par In this way, we solve iteratively for the $\omega_k$ coefficients of Eq. (\ref{eq:omegaseries}), at increasing orders of $k$, to generate a $L$-dependent expression for the QNF. The output of this method is most reliable in the large-$\ell$ regime, and for small values of $Q$ and $q$. Outside of these limitations, the method is fairly robust: while a number of numerical and semi-classical techniques require input values for space-time and field parameters from the onset of the computation (e.g. consider Refs. \cite{Konoplya2019_recipes,Dias2009_Paraspectral1,Dias2009_Paraspectral2}), here we maintain the masses, charges, overtones, and harmonics as free variables. It is for this reason that we employ this method here: to provide insight into the behaviour of the QNFs throughout the phase space, in order to identify the parameters that warrant further exploration.

\par As we shall discuss in the next section, the QNF frequency spectrum associated with the RNdS black hole space-time and a massive charged test field is complicated by the number of free parameters and the breaking of the QNF reflection symmetry. For these reasons, it is useful to employ a flexible technique like this WKB-based method to scan the parameter space for general trends in the QNF behaviour and to identify key benchmark points that warrant further investigation (which we reserve for a follow-up work).

\subsubsection{Classifying charged QNMs in the RNdS space-time: \label{subsubsec:classQNM}}

\par {\color{black}{Within spherically-symmetric black holes with $\Lambda>0$, Ref. \cite{Konoplya2024_TwoRegimes} asserts that the QNMs can be categorised into two main branches:
\begin{outline}
\1[$(I)$] \enquote{Schwarzschild modes}: the modes of an asymptotically-flat black hole corrected by a $\Lambda >0$ term;
\1[$(II)$] \enquote{de Sitter modes}: the modes of an empty de Sitter space-time corrected by the presence of a black hole.
\end{outline}
\noindent When $\Lambda \rightarrow 0$ and $M \rightarrow 0$, these reduce to Schwarzschild and pure de Sitter modes, respectively. While there are both real and imaginary components for large-$\mu M$, the real modes vanish for smaller values of $\mu M$ for the de Sitter branch \cite{Konoplya2024_TwoRegimes}.}}

\par As first stipulated in Ref. \cite{Cardoso2017_QNMsSCC}, and further explored in Refs. \cite{DiasReallSantos2018_SCC,
Mo2018_SCC,
DiasSantos2020_RNdSinstability,
Papantonopoulos2022}, QNMs in the RNdS space-time can be classified into three qualitatively distinct types based on the structure of their QNF solution: a $(i)$ \enquote{photon-sphere} type, a $(ii)$ \enquote{de Sitter} type, and a $(iii)$ \enquote{near-extremal} type. The photon-sphere type connects smoothly to the \enquote{Schwarzschild} branch, while the de Sitter type connects to the \enquote{de Sitter} branch \cite{Cardoso2018_QNMsSCC}. The near-extremal type, on the other hand, is unique to the RNdS space-time. We find that each type dominates a particular region within the phase space sketched in Fig. \ref{fig:sharkfin}. In keeping with the conventions of Ref. \cite{Cardoso2017_QNMsSCC}, we note the following QNF behaviours:

\begin{outline}
\1[$(i)$] photon-sphere modes lie in the region beneath the $M=Q$ line and approaching Line $NU$; they are characterised by large $\mathfrak{R}e \{ \omega \}$, where their $\mathfrak{I}m \{ \omega \}$ is related to the instability time scale of null geodesics near the black hole photon sphere in the large-$\ell$ regime. Closely following along Line  $NU$, the charged Nariai branch, and particularly for smaller $Q$ values,
\begin{eqnarray}
    \mathfrak{I}m \{ \omega_{PS} \} \approx -i \left(n + \frac{1}{2} \right) \kappa_+ \;;
\end{eqnarray}
\1[$(ii)$] de Sitter modes can be found in the region near Point $O$, following closely along 
Line $OU$ and in competition with the near-extremal modes. Here, $\kappa_c \sim 1/L_{dS}$ and
\begin{equation}
\omega_{dS_{n=0}} \approx -i \ell  \kappa_c \;, \quad \omega_{dS_{n \neq 0}} \approx -i (\ell + n + 1) \kappa_c \;;
\end{equation}
\1[$(i)$] near-extremal modes arise near the line $OU$ where $r_- \sim r_+$, 
\begin{equation}
\omega_{NE} \approx - i (\ell + n + 1) \kappa_- = - i (\ell + n + 1) \kappa_+ \;.
\end{equation}
\end{outline} 
\noindent These hold for electrically neutral and charged scalar test fields. We note that we observe a non-zero $\mathfrak{R}e \{ \omega \}$ part in each of these regions for low values of $\mu$ and $q$. However, in the case of the near-Nariai region, this contribution is very small: $\mathfrak{R}e \{ \omega \} \sim \mathcal{O}(0.01)$ for $Q<0.1$.

\par Let us now consider the QNM spectrum itself. Irrespective of the space-time parameters, when  studying $q=0$ QNM problems in spherically symmetric space-times, there are two sets of QNFs for which the imaginary parts are identical and the real parts are of equal magnitude but opposite sign. This is a natural consequence of the QNM reflection symmetry $\omega \rightarrow - \omega^*$, such that the complex conjugate $\omega^*$ is the QNF corresponding to the QNM $\varphi^*$ that satisfies Eq. (\ref{eq:odefullRNdS}). Upon introducing a non-zero field charge, the QNM reflection symmetry is broken, leaving us with two distinct sets of QNF solutions, {\color{black}{such that if $\varphi$ has a charge $q$, then $\varphi^*$ has a charge $-q$. To elaborate: if $\omega = \omega_a + i \omega_b$ is a QNF associated with the QNM $\varphi$, then $-\omega^*= -\omega_a + i \omega_b$ is a QNF associated with the QNM $\varphi^*$. In other words, we must allow for both positive and negative values of $\mathfrak{R}e \{ \omega \}$. We assume $q >0$. 

\par Furthermore, we find that our WKB-based analysis yields two families of QNFs for nonzero $q$, as observed in Refs. \cite{DiasReallSantos2018_SCC,Papantonopoulos2022}.}} Following the convention of Ref. \cite{DiasReallSantos2018_SCC}, we refer to the two families of solutions as a \enquote{black-hole family} $\omega_+$ and a \enquote{cosmological horizon family} $\omega_c$. For a fixed $q$, we find that $\vert \mathfrak{Re} \{ \omega_+ \} \vert~<~\vert \mathfrak{Re} \{ \omega_c \} \vert$ and $\vert \mathfrak{Im} \{ \omega_+ \} \vert~<~\vert \mathfrak{Im} \{ \omega_c \} \vert$. 

\par In the next section, we shall focus on the influence of $\mu$ on the QNF spectrum. We highlight the anomalous decay first noted in Ref. \cite{Lagos2020_Anomalous} for the Schwarzschild case, in which the QNF damping decreases with the angular momentum number for $\mu$ below a critical field mass $\mu_{crit}$. 

\subsubsection{On the scalar field mass of the QNF: \label{sec:critmass}}

\begin{figure}[t]
\centering
\includegraphics[width=.6\columnwidth]{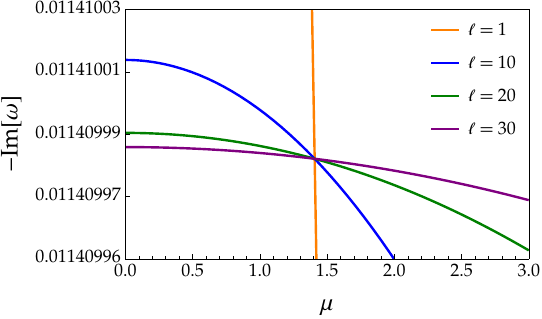}
\caption{\label{fig:ImwVSmuNariai} \textit{A graphical indication of the critical mass $\mu_{cit}\sim  1.4083$ at Point $N$, where $r_+ \sim r_c$ for $L^2_{dS}=1$, $M=1/\sqrt{27}$, and $Q=0$.}}
\end{figure}
\par As discussed in Section \ref{chap:intro} for the scalar QNFs within a Schwarzschild black hole space-time, an increase in $\ell$ or an increase in $\mu$ leads to increased $\vert \mathfrak{Re} \{ \omega \} \vert$. Similarly, $\vert \mathfrak{Im} \{ \omega \} \vert$ scales with $\ell$. On the other hand, $\vert \mathfrak{Im} \{ \omega \} \vert$ decays with increasing scalar field mass $\mu$. As such, heavier modes are expected to be less damped and longer lived. This is demonstrated in Fig. \ref{fig:ImwVSmuNariai}, where we plot the imaginary part of the QNF for increasing values of $\mu$. There, we set $M=1/\sqrt{27}$ and $Q=0$, corresponding to Point $N$ of Fig. \ref{fig:sharkfin}: the uncharged Nariai solution at which $r_+ \sim r_c$. Irrespective of the value of $\ell$, the magnitude of the imaginary part of the QNF $\vert \mathfrak{I}m \{ \omega \} \vert$ decreases with $\mu$.

\par For (massive) scalar QNMs in a spherically symmetric space-time, it is expected that $\vert \mathfrak{Im} \{ \omega \} \vert$ scales with $\ell$ \cite{refBertiCardoso,refKonoplyaZhidenkoReview}. However, this relationship was only observed for heavier modes in the Schwarzschild case (e.g. Fig. 3 of Ref. \cite{Lagos2020_Anomalous}). Similarly, we do not see this throughout Fig. \ref{fig:ImwVSmuNariai}: $\vert \mathfrak{Im} \{\omega_{\ell=10} \} \vert > \vert \mathfrak{Im} \{\omega_{\ell=30} \} \vert$ for lower values of $\mu$ and $\vert \mathfrak{Im} \{\omega_{\ell=10} \} \vert < \vert \mathfrak{Im} \{\omega_{\ell=30} \} \vert$ for larger values of $\mu$. In other words, only past a certain mass value do we observe \enquote{regular} QNF behaviour. In Fig. \ref{fig:ImwVSmuNariai}, this \enquote{critical mass} is given by $\mu_{crit} \sim 1.4083$ and corresponds to a very small $\mathfrak{I}m \{ \omega \} \sim -0.0114$. We observe that for these values of $\mu=\mu_{crit}$ and $(M,Q)=(1/\sqrt{27},0)$, $M \mu \sim 0.3$; this is the smallest possible $\mu_{crit}$ for the RNdS black hole parameterised as $L_{dS}=1$.

\par This \enquote{critical mass}  $\mu_{crit}$ value has been observed in Schwarzschild (and Kerr) black hole space-times \cite{Lagos2020_Anomalous}, as well as for the charged, massive scalar field in the RN and RNdS space-times \cite{Fontana2020_RNanomalous,Papantonopoulos2022}. Here, we study this effect in greater detail, to determine the parameter range for which it is valid.

\par We can solve for $\mu_{crit}$ by setting $\mathfrak{Im} \{ \omega_{-2} \} =0$ \cite{Papantonopoulos2022} (see \ref{app:num}). At the lowest order in $Q$, 
\begin{equation} \label{eq:mucritLO}
   \mu_{crit}^2 =  \frac{18045 \Lambda  M^2+137}{29160 M^2} \;.
\end{equation}
As shown in Fig. \ref{fig:ImwVSmuNariai}, we can also determine $\mu_{crit}$ graphically by plotting $-\mathfrak{Im} (\omega)$ vs $\mu$. A common point of intersection, irrespective of the hierarchy in $\ell$, denotes the value of $\mu_{crit}$. We observe that this coalescence indicates a negligible dependence of the QNF on $\ell$ at $\mu=\mu_{crit}$, and confirm it analytically.

\par In particular, we observe in Fig. \ref{fig:ImwVSmuNariai} that for the uncharged Nariai case, the critical mass corresponds to $\mu_{crit} \sim 1.4083$ for all $q$. This same result is found using Eq. (\ref{eq:mucritLO}) for $q=0$. This is the maximum $M$ for which we see an intersection. We do not observe further intersections along the Line $NU$, and since $Q=0$, the charge $q$ has no influence on the QNF at Point $N$ (recall that $q$ and $Q$ couple in the potential, Eq. (\ref{eq:odefullRNdS})). For $q \approx 0$, the intersection point shifts to the right of Fig. \ref{fig:sharkfin} as we increase $Q$, and as we increase $M$ (from zero) the intersection point shifts left. For fixed values of $M$ and $Q$, the intersection point shifts left as we increase $q$ (along the small domain of $q$). 

\par When $q=0$, the influence of $Q$ is negligible on $\mu_{crit}.$ We find that large $M$ corresponds to a small $\mu_{crit}$; $\mu_{crit}$ decreases with increasing $M$, such that for $M \sim 0$, $\mu_{crit} \sim 20$. When $M > 1/\sqrt{9 \Lambda}$ for non-zero $Q$, we observe growing rather than decaying modes for large values of $\mu$ e.g. when $\mu > 5$ for $\ell \sim 10$ and $\mu > 15$ for $\ell \sim 30$. For $q>0,$ we do not observe an intersection of all lines, but the anomalous behaviour in which $\vert \mathfrak{Im} \{\omega_{\ell} \} \vert > \vert \mathfrak{Im} \{\omega_{\ell+1} \} \vert$ is noted for smaller values of $\mu$ (see Fig. \ref{fig:ImwVSmu_MQc01}). As we would expect from the coupling between $q$ and $Q$, $Q$ has a more obvious effect for the nonzero $q$, such that the spacing between modes increases with $Q$.   

\begin{figure}[t]
        \centering
            \includegraphics[width=.6\textwidth]{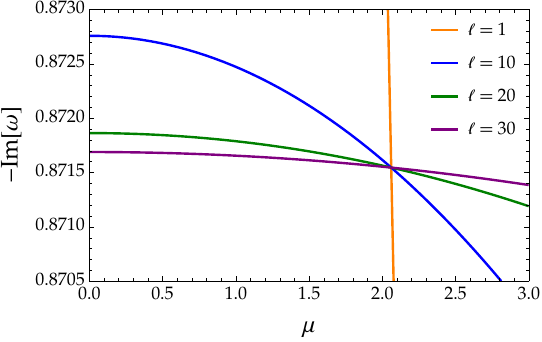}
        \caption{\label{fig:ImwVSmu_MQ}
        \textit{At Point $E$, where $M=Q=0.104$ for $L^2_{dS}=1$, the critical mass is $\mu_{crit} \sim  2.065$ for $q=0$.}}   
        \end{figure}
    \begin{figure}[t]
        \centering
        \includegraphics[width=.6\textwidth]{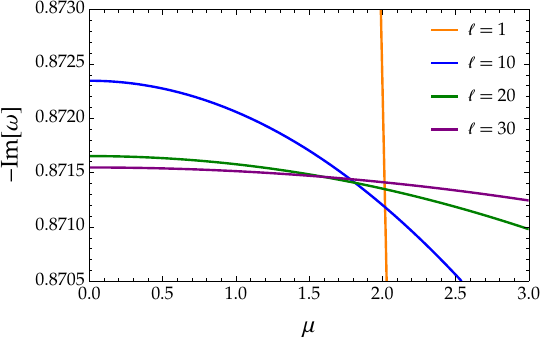}
        \caption{\label{fig:ImwVSmu_MQc01} \textit{At Point $E$, where $M=Q=0.104$ for $L^2_{dS}=1$, there is no longer an intersection to denote $\mu_{crit}$ for $q=0.1$.}}    
        \end{figure}

\subsubsection{QNFs and the Festina-Lente bound: \label{sec:ConnectionGW}}

\par As a final comment and brief aside on the scalar field mass, we consider whether scalar QNFs with a non-zero mass and charge satisfying the Festina-Lente bound could be observable. When the Compton wavelength of the massive bosonic field is of the order of the black hole’s radius, the scaling for the suppression of the instability timescale is governed by $M \mu~\lesssim~1$ \cite{refBertiCardoso,refKonoplyaZhidenkoReview}.  In the case of ultralight scalar fields, this instability timescale is of the order of seconds for black holes of mass $M \sim M_{\odot}$, or hundreds of years for the heavier $M \sim 10^9 M_{\odot}$ black holes  \cite{refKonoplyaZhidenkoReview,BritoCardosoPani2020_Superradiance}. As such, if we are to observe QNMs, we can consider $M \mu \sim \mathcal{O}(1)$. As we demonstrated in Ref. \cite{Chrysostomou2023_EPJC}, we can then utilise dimensional analysis to isolate a mass range for the scalar field. Using SI units $M = m_{\rm BH} G /c^2$ and $\mu = mc/\hbar$, we can express the mass of a bosonic field as
\begin{equation} \label{eq:massbound}
    m = \frac{1}{m_{\rm BH}} \frac{\hbar c}{G} M\mu \;.
\end{equation}
\noindent Since $\hbar c /G \sim 10^{-16} $ kg$^2$ and $1 M_{\odot} \sim 10^{30}$ kg, we can scale the black hole mass as $m_{\rm BH} \sim 10^\chi M_{\odot}$. Returning to natural units with $c=1$ (c.f. Eq. (4.25) of Ref. \cite{BritoCardosoPani2020_Superradiance}), we then have
\begin{equation} \label{eq:Mm}
m \sim 10^{-( \chi + 10)}M \mu \, \rm{ eV} \;.
\end{equation}

\par Eq. (\ref{eq:Mm}) serves as a rough detectability bound on the sensitivity of the QNF spectrum corresponding to an astrophysical black hole to ultralight scalar fields. We have seen that for the critical mass in the regime $\mu M \gg q Q$, $\mu_{crit} M \leq 0.3$.  To satisfy the Festina-Lente bound of $m > 10^{-3}$ eV, this means we can only consider compact objects corresponding to $\chi < -8$, such as micro black holes. With this, we surmise that the Festina-Lente bound rules out the possibility of observing weakly charged scalar QNMs from astrophysical black holes.

%%%%%
%CONCLUIONS
%%%%%

\section{Conclusions}
\par We began this work with an extensive review of the literature, in order to consolidate known but scattered results into a single manuscript and to stress the importance of beginning QNM studies with quantitative contextualisation of the phase space. In the RNdS context, we must take into account the effects of the non-zero cosmological constant on the black hole solution space. These include:
\begin{itemize}
    \item[$(i)$] the emergence of the $Q~>~M$ region of the parameter space, allowing for non-superextremal black holes whose charge exceed their mass, as well as an analytically-derived upper bound on the black hole mass;
    \item[$(ii)$] the existence of a third horizon, the cosmological horizon, that enforces an upper bound on the size of the black hole,  $r_+ \leq r_c \leq L_{dS}$;
    \item[$(iii)$] limiting cases such as the near-Nariai solution that corresponds to a black hole whose event horizon and cosmological horizon are infinitesimally close but do not actually coincide.
\end{itemize}

\par The focus of this study was on the interplay of black hole and scalar field parameters. The influence of the scalar field mass on the QNF spectrum was of particular interest, where we analysed the properties of $\mu = \mu_{crit}$, beneath which $\mathcal{I}m \{ \omega \}$ scales inversely with $\ell$. For $L^2_{dS} = 3/\Lambda = 1$, we have determined the maximum black hole mass for which $\mu_{crit}$ exists in the case of massive charged scalar QNMs: at Point $N$ corresponding to the uncharged Nariai solution. We note with interest that black hole mass scales inversely with $\mu_{crit}$: for $M \sim 0$, $\mu_{crit} \sim 20$ whereas for the maximum $M \sim 1/\sqrt{27}$, we obtained the minimum $\mu_{crit} \sim 1.4$. This demonstrates a compensatory behaviour between scalar field mass and cosmological constant. While it is known that the cosmological constant has a suppressive effect on the QNF spectrum, our results suggest that \enquote{regular} QNF behaviour only takes place if the scalar field mass is sufficiently large with respect to the cosmological constant. Furthermore, for vanishing $q$ and low $Q$, the dimensionless constant is $\mu M < 0.3$ for $\mu = \mu_{crit}$.

\par For fixed values of $M$ and $Q$, the value of $\mu_{crit}$ decreases as we increase $q$. When $q=0$, the influence of $Q$ on $\mu_{crit}$ is negligible. For larger values of $q>1$, we did not observe an intersection in the $-\mathfrak{Im} \{ \omega \}$ vs $\mu$ plot, but the anomalous behaviour in which $\vert \mathfrak{Im} \{\omega_{\ell} \} \vert > \vert \mathfrak{Im} \{\omega_{\ell+1} \} \vert$ is noted when $\mu>0$ is sufficiently small. Due to the coupling between the field charge $q$ and the black hole charge $Q$, $Q$ has a more obvious effect for a nonvanishing $q$, such that the spacing between individual modes increases with $Q$ (compare Fig. \ref{fig:ImwVSmuNariai} at Point $N$ with Figs. \ref{fig:ImwVSmu_MQ} and \ref{fig:ImwVSmu_MQc01} at Point $E$). We emphasise that this value of $\mu_{crit}$ corresponds to a point in the QNF spectrum for which dependence on the angular momentum number $\ell$ is negligible. Moreover, it is interesting to note that the introduction of the scalar field charge offsets this value (see Figs. \ref{fig:ImwVSmu_MQ} and \ref{fig:ImwVSmu_MQc01}), suggesting perhaps that the behaviour of the charged scalar field cannot be decoupled from the angular momentum of the QNF.

\par Furthermore, using dimensional analysis, we also found evidence suggesting that the Festina-Lente bound rules out the possibility of observing weakly charged scalar QNFs from astrophysical black holes. For scalar test fields oscillating within the exterior of an astrophysical black hole, recent studies \cite{BritoCardosoPani2020_Superradiance,BertiBrito2019_UltraLight1GW,BertiBrito2017a_UltraLightStochastic,BertiBrito2017b_UltraLightLIGOLISA} indicate that modes scaling as $M\mu \sim \mathcal{O}(1)$ may undergo a superradiant amplification on timescales of suitable length to be observed by current and next-generation GW detectors. Here, we find that $M \mu_{crit} \sim 0.3$. By Eq. (\ref{eq:Mm}), we infer that only from compact objects of the order $m \lesssim 10^{-8} \; M_{\odot}$ can we expect to observe QNMs satisfying the Festina-Lente bound.

\par From our analysis of the evolution of the potential within the phase space, we found that a barrier potential characteristic of QNM behaviour was shown to exist on $r_+ \leq r \leq r_c$ for $\ell >0$ (with the \enquote{valley} suggestive of superradiant amplification suppressed or lying beyond $r = r_c$). The only exception to this could be found in extremised regions of the RNdS parameter space, where the coalescence of horizons suppressed the potential (i.e. Figs. \ref{fig:RNdSpotential-chNariai} and \ref{fig:RNdSpotential-cold}) and the $\ell=0$ case (see Fig. \ref{fig:potential-zeroL}, where the \enquote{valley} falls within $r_+ < r < r_c$). However, as demonstrated in Fig. \ref{fig:RNdSpotentials2muq}, increasing $\mu$ can uplift the local maximum of a suppressed potential within non-extremised regimes. This suggests that WKB-based techniques can be applied to these regimes. Additional support for this assumption can be found in the recent work, Ref. \cite{Konoplya2024_TwoRegimes}, which suggested that applying the WKB method for large $\mu M$ will yield sufficiently accurate results, satisfying the criteria of Ref. \cite{Konoplya2019_recipes} for accurate WKB computations.

%%%%%
\subsection*{Acknowledgements}
AC acknowledges the support provided during the course of this project by the National Research Foundation (NRF) of South Africa and the Department of Science and Innovation through the SA-CERN programme, as well as that of a Campus France scholarship and a research grant from the L’Oréal-UNESCO's \textit{For Women in Science} Programme. AC is currently supported by the Initiative Physique des Infinis (IPI), a research training program of the Idex SUPER at Sorbonne Universit\'e. ASC is supported in part by the NRF. SCP is supported by the National Research Foundation of Korea (NRF) grant funded by the Korea government (MSIT) RS-2023-00283129 and RS-2024-00340153. The authors thank Hajar Noshad for discussions and collaboration during the initial stages of this project. We also thank Roman Konoplya for his indispensable comments and advice.

\appendix

\renewcommand{\theequation}{A.\arabic{equation}}
\section{The Nariai black hole solution \label{app:Nariai}}

\par Line $NU$ of Fig. \ref{fig:sharkfin} is referred to in the text and literature as the \enquote{charged Nariai branch}. While we loosely refer to it in the diagram as the solution for which $r_+ = r_c$, the horizons do not actually meet but instead become infinitesimally close. This is a subtlety not often addressed in the literature. Furthermore, due to its relevance to the Festina-Lente bound, we discuss this Nariai solution explicitly and how it relates to black holes in de Sitter space-time. The (uncharged) Nariai solution was first constructed in 1950 \cite{Nariai1950_StaticSol,Nariai1951_NewSol}, from the isotropic form of the line element,
\begin{equation}
ds^2 = -e^{\nu(r)} dt^2 + e^{\mu(r)} (dr^2 + r^2 d\theta^2 + r^2 \sin^2 \theta d \phi^2 ) \;.
\end{equation}
\noindent Assuming a homogenous static universe with spherical symmetry, Nariai determined that
{\setlength{\mathindent}{0pt}
\begin{equation} \label{eq:Nariai1}
ds^2 = \frac{1}{\Lambda} \left[ -(X \cos \{\log r\} + Y \sin \{\log r \})^2 dt^2 + \frac{1}{r^2} (dr^2 + r^2 d\theta^2 + r^2 \sin^2 \theta d\phi^2 ) \right] 
\end{equation}
}
\noindent for arbitrary constants $X$ and $Y$, satisfied the Einstein field equations for an empty universe with a non-zero cosmological constant. We can reformulate this into a more tractable expression with the introduction of a few transformations \cite{Nariai1951_NewSol}, 
\begin{eqnarray}
t &=& \tau \left( \frac{\Lambda}{X^2 + Y^2} \right)^{1/2} \;, \nonumber \\
r &=&  \exp \bigg \{ \pm \chi + \tan^{-1} \Bigg \{\frac{Y}{X} \Bigg \} \bigg \} \;, \\
r_1 &=&  L \sin \{ \chi \} \;, \nonumber
\end{eqnarray}
\noindent where $L^2=1 / \Lambda$. Upon making these transformations in Eq. (\ref{eq:Nariai1}), we obtain 
\begin{equation}
ds^2 = -\cos^2 \chi d\tau^2 + L^2 (d\chi^2 + d\Omega^2) \;,
\end{equation}
\noindent for $d\Omega^2 = d\theta^2 + \sin^2 \theta d\phi^2$. Since $\sin \chi = r_1/L$, we can rewrite this expression as
\begin{equation} \label{eq:Nariai2}
ds^2 = -\left( 1 - \frac{r_1^2}{L^2} \right) d\tau^2 + \left( 1 - \frac{r_1^2}{L^2} \right)^{-1} dr_1^2 + L^2 d\Omega^2 \;.
\end{equation}
\noindent As an aside, note that we can compare this to the purely de Sitter solution, with $L^2_{dS}=3/\Lambda$,
\begin{eqnarray}
ds^2 &=& -\cos^2 \chi d\tau^2 + L_{dS}^2 (d\chi^2 + \sin^2 \chi d\Omega^2) \;,
 \\
&=& -\left( 1 - \frac{r_1^2}{L_{dS}^2} \right) d\tau^2 + \left( 1 - \frac{r_1^2}{L_{dS}^2} \right)^{-1} dr_1^2 + r_1^2 d\Omega^2 \;. 
\end{eqnarray}
\noindent For the purely radial case $(d\Omega^2=0)$, Eq. (\ref{eq:Nariai2}) is nearly identical to pure de Sitter space. Observe that the Nariai space-time is spherically-symmetric, homogeneous and locally static. However, it is not isotropic nor is it globally static. It has the geometry $dS_2 \times \mathbb{S}^2$ and a topology $\mathbb{R}\times\mathbb{S}^1\times\mathbb{S}^2$. The space-time satisfies $R_{\mu \nu} = \Lambda g_{\mu \nu}$, where $\Lambda=1/L^2$, and has constant Ricci scalar curvature, $R = 4\Lambda$ (see also Refs. \cite{Bousso1996_Nariai,
Cardoso2003_ExtremalSchw,
refCardosoDiasLemos2004_Nariai-BR-antiNariai,
CasalsDolan2009_Nariai} for further discussion).
 
\par Under a particular limiting procedure, Ginsparg and Perry \cite{GinspargPerry1982_NariaidS} showed that the Nariai solution can be
generated from the limiting case in which the event horizon and the cosmological horizon of the Schwarzschild de Sitter space-time approach one another. That is, the extremal Schwarzschild de Sitter black hole, where $M = 1/\sqrt{9 \Lambda}$, sometimes referred to as the \enquote{Nariai limit}. Analogously, Hawking and Ross \cite{HawkingRoss1995_NariaiRNdS} obtained the charged Nariai solution of Bertotti \cite{Bertotti1959_BRsol} and Robinson \cite{Robinson1959_BRsol} as a limiting case of the RNdS black hole.

\par Along the $NU$ branch of Fig. \ref{fig:sharkfin}, the proper distance between the the event and the cosmological horizons remains finite in the regime where $r_+ \rightarrow r_c$, such that $r_+ \rightarrow \varrho - \epsilon$ and $r_c \rightarrow \varrho + \epsilon$, for some infinitesimally small $\epsilon$ \cite{Romans1991_ColdLukewarmRNdS,
HawkingRoss1995_NariaiRNdS,
Anninos2012_Musings}. This is best illustrated by a change in coordinates, where we use the example from Ref. \cite{vanRiet2019_FLevapBHdS} that allows for a smooth transition of the black hole from a regular to extremal state. There, the coordinate $r=r_g$ is introduced, at which the competition between the gravitational attraction of the black hole and the accelerating expansion of the universe cancel and $f(r_g)=f'(r_g)=0$. A \enquote{geodesic observer} situated at this point travels along a time-like Killing vector field which is also a geodesic. Suppose we let
\begin{equation}
\rho \rightarrow \frac{r-r_g}{\sqrt{ \vert f(r_g) \vert }} \;, \quad \tau \rightarrow \sqrt{ \vert f(r_g) \vert } \;t \;.
\end{equation}  
\noindent The metric then becomes
\begin{equation}
ds^2 = - \frac{f(r)}{\sqrt{ \vert f(r_g) \vert }} d\tau^2 + \frac{\sqrt{ \vert f(r_g) \vert }}{f(r)} d\rho^2 + r^2 d\Omega^2 \;,
\end{equation}
\noindent and the magnitude of the electric field is unchanged.  

\par From the perspective of the geodesic observer, these two horizons become infinitesimally close along the $NU$ branch, but they do not coincide. There,
\begin{equation}
\frac{U(r)}{U(r_g)} \rightarrow 1-\frac{\rho^2}{L_{dS_2}^2}\;, \quad r^2 \rightarrow r_c^2 \;.
\end{equation}
\noindent This produces the $dS_2 \times S^2$ metric,
\begin{equation}
ds^2 = - \left( 1-\frac{\rho^2}{L_{dS_2}^2}\right)d\tau^2 + \left( 1-\frac{\rho^2}{L_{dS_2}^2} \right)^{-1} d \rho^2 + r_c^2 d \Omega^2 \;, 
\end{equation} where
\begin{equation}
L_{dS_2}^2 = \frac{2}{f''(r_c)} = \frac{1}{6} \left( \frac{1}{\sqrt{1-12Q^2}} + 1 \right) = \left(3 - \frac{Q^2}{r_c^4} \right)^{-1} \;.
\end{equation}

\noindent The corresponding $S^2$ radius on the $NU$ branch is 
\begin{equation}
r_c(Q) = \sqrt{ \frac{1}{6} \left( 1 + \sqrt{1-12Q^2} \right)} \;.
\end{equation}
\noindent We can see that this is equivalent to the value of the cosmological horizon $r_c$ in Eq. (\ref{eq:radii}).

\renewcommand{\theequation}{B.\arabic{equation}}
\section{Details on the semi-classical method \label{app:num}}

\par When considering massive charged QNMs oscillating on the RNdS space-time, there are a number of free parameters that must be taken into account for the scalar QNF, $(\mu,q,\ell,n)$, and the black hole space-time, $(M,Q,\Lambda)$. To calculate the QNFs, we choose to make use of the semi-classical modified WKB method \cite{refBHWKB0,refBHWKB0.5,refBHWKB1,Konoplya2003,
Konoplya2019_recipes}, as outlined in Ref. \cite{Papantonopoulos2022}, such that the QNF is generated in the form of a series expansion in $L$. With this technique, the black hole and scalar field input variables are left as free parameters (i.e. charges, masses, etc. do not need to be defined from the onset of the computation, as is the case in a variety of other methods \cite{refKonoplyaZhidenkoReview,Dias2015_Num}). In this way, values for the black hole and scalar field input parameters can be introduced after the iterative procedure has been applied, therefore allowing for a complete scan of the available phase space for a number of different combinations of variables.

\par For $N=n+1/2$ and the derivative being $V^j$ taken with respect to a generalised tortoise coordinate, we compute the values of the QNF series terms using Eqs. (\ref{eq:BorelQNF}-\ref{eq:Vj}) \cite{Hatsuda2019_WKB,Papantonopoulos2022}, for which
{\setlength{\mathindent}{4pt}
\begin{eqnarray}
 U && =  N \sqrt{ \frac{-V^{(2)}}{2}} + \frac{i}{64} \left[- \frac{1}{9} \left( \frac{V^{(3)}}{V^{(2)}} \right)^2 (7 + 60N^2) + \frac{V^{(4)}}{V^{(2)}}(1+4N^2) \right] \nonumber \\
&& \quad+ \frac{N}{2^{3/2} 288} \Bigg[ \frac{5}{24} \left( \frac{(V^{(3)})^4}{(-V^{(2)})^{9/2}} \right)(77+188N^2) + \frac{3}{4} \left( \frac{(V^{(3)})^2 V^{(4)}}{(-V^{(2)})^{7/2}} \right) (51+100N^2)  \nonumber \\
&&\quad + \frac{1}{8} \left( \frac{(V^{(4)})^2}{(-V^{(2)})^{5/2}} \right) (67+68N^2) + \left( \frac{V^{(3)} V^{(5)}}{(-V^{(2)})^{5/2}} \right) (19+28 N^2) \nonumber \\
&& \quad +\left( \frac{V^{(6)}}{(-V^{(2)})^{3/2}} \right) (5+ 4N^2) \Bigg] \;,  
\end{eqnarray}
}
\noindent as first derived in Eq. (1.5b) of Ref. \cite{refBHWKB1}. The coefficients featured in the series expansion Eq. (\ref{eq:rmax}) are
{\setlength{\mathindent}{0pt}
\begin{eqnarray}
r_0 &&\approx   3M-\frac{2Q^2}{3M} - \frac{4 Q^4}{27 M^3} + ...\;, \\
 r_1 &&\approx 27 \Lambda  M^5 \left(2 \Lambda -3 \mu ^2\right)-3 M^3 \left(\Lambda -3 \mu ^2\right)-\frac{M}{3}  -9 \left(q M^2 \omega \right) Q \nonumber \\
 && \quad + \Bigg [M \left(3 q^2+\frac{2 \Lambda }{3}-5 \mu ^2\right)+24 \Lambda  M^3 \left(3 \mu ^2-2 \Lambda \right)+\frac{5}{27 M}\Bigg]Q^2 \nonumber \\
 && \quad +2 q Q^3 \omega +Q^4 \left[4 \Lambda  M \left(2 \Lambda -3 \mu ^2\right)-\frac{14}{81 M^3}\right] + ... \;.
\end{eqnarray}
}
\noindent Similarly, the contributions to the series expansion of the potential, Eq. (\ref{eq:Vmax}), are
{\setlength{\mathindent}{0pt}
\begin{eqnarray}
\nonumber
 V_0 &\approx &  \frac{1}{27 M^2} \left(1-9 \Lambda M^2 \right) +\frac{Q^2}{81 M^4} +\frac{4 Q^4}{729 M^6} + ... \\
\nonumber
 V_1 &\approx & -\frac{4 \Lambda }{9}+\frac{\mu ^2}{3}+2 \Lambda ^2 M^2-3 \Lambda  \mu ^2 M^2+\frac{2}{81 M^2} + \frac{2 qQ \omega }{3 M} \nonumber \\
 &&\quad +\Bigg [ -\frac{q^2}{9 M^2}-\frac{8 \Lambda ^2}{9}+\frac{4 \Lambda  \mu ^2}{3} +\frac{4}{729 M^4}+\frac{4 \Lambda }{81 M^2}-\frac{\mu ^2}{27 M^2} \Bigg] Q^2 \nonumber \\
 &&\quad + \frac{4 q \omega }{27 M^3} Q^3+ \Bigg[ -\frac{4 q^2}{81 M^4}+\frac{10}{6561 M^6}+\frac{16 \Lambda }{729 M^4} -\frac{4 \mu ^2}{243 M^4} \nonumber \\
 &&\quad\quad -\frac{8 \Lambda ^2}{81 M^2}+\frac{4 \Lambda  \mu ^2}{27 M^2}\Bigg]Q^4 + ...
\end{eqnarray}
}
\noindent The first term in the QNF series expansion is given by
{\setlength{\mathindent}{1cm}
\begin{eqnarray}
    \omega_0 &\approx & \frac{q Q}{3 M} -\frac{i \sqrt{\frac{1}{3}-3 \Lambda  M^2}}{6 M}  +\frac{i Q^2 \sqrt{\frac{1}{3}-3 \Lambda  M^2} \left(18 \Lambda  M^2+1\right)}{108 M^3 \left(9 \Lambda  M^2-1\right)} +\frac{2 q Q^3}{27 M^3}\nonumber \\
    &&\quad  +\frac{i Q^4 \left(1-18 \Lambda  M^2\right)^2}{432 \sqrt{3} M^5 \left(1-9 \Lambda  M^2\right)^{3/2}} +...
\end{eqnarray}
}

\noindent To solve for the critical mass, we set $\mathfrak{Im} \{ \omega_{-2} \} =0$ \cite{Papantonopoulos2022}. 

\par As already stated, the value of this method lies in its flexibility: we were able to explore the full parameter space of the RNdS black hole at low computational cost, in order to identify areas of interest for subsequent study. However, we must address a number of weaknesses in this method for our particular QNM analyses. As already mentioned, the method itself does not yield very precise results. The Borel-resummation technique of Ref. \cite{Hatsuda2019_WKB} derives the function $U$ in a simplified manner, which is equivalent to the result of Ref. \cite{refBHWKB1}; the higher-order corrections to Ref. \cite{refBHWKB1}, summarised in Ref. \cite{Konoplya2019_recipes}, are not included. Furthermore, the WKB method is most accurate in the eikonal regime. When $\ell \rightarrow \infty$, $\mathfrak{I}m \{ \omega \}$ is independent of $\ell$. Recall that in asymptotically flat and de Sitter space-times, the imaginary component tends towards the Lyapunov constant describing the decay time scale of the perturbations \cite{refGoebel1972}. Techniques that capture the behaviour of the imaginary component of the QNF more precisely are better suited to investigations concerning the imaginary component. 
%
%\section*{References}
\bibliographystyle{iopart-num} %utphys
\bibliography{zotero}% Produces the bibliography via BibTeX.

\providecommand{\newblock}{}
\begin{thebibliography}{10}
\expandafter\ifx\csname url\endcsname\relax
  \def\url#1{{\tt #1}}\fi
\expandafter\ifx\csname urlprefix\endcsname\relax\def\urlprefix{URL }\fi
\providecommand{\eprint}[2][]{\url{#2}}

\bibitem{Israel1968_NoHair-RN}
Israel W 1968 {\em Commun. Math. Phys.\/} {\bf 8} 245--260

\bibitem{Bousso1996_Nariai}
Bousso R 1997 {\em Phys. Rev. D\/} {\bf 55} 3614--3621 (\textit{Preprint}
  \eprint{gr-qc/9608053})

\bibitem{vanRiet2019_FLevapBHdS}
Montero M, Van~Riet T and Venken G 2020 {\em JHEP\/} {\bf 01} 039
  (\textit{Preprint} \eprint{1910.01648})

\bibitem{vanRiet2021_FL}
Montero M, Vafa C, Van~Riet T and Venken G 2021 {\em JHEP\/} {\bf 10} 009
  (\textit{Preprint} \eprint{2106.07650})

\bibitem{refBertiCardoso}
Berti E, Cardoso V and Starinets A~O 2009 {\em Class. Quant. Grav.\/} {\bf 26}
  163001 (\textit{Preprint} \eprint{0905.2975})

\bibitem{refKonoplyaZhidenkoReview}
Konoplya R~A and Zhidenko A 2011 {\em Rev. Mod. Phys.\/} {\bf 83} 793--836
  (\textit{Preprint} \eprint{1102.4014})

\bibitem{SimoneWill1991_MassiveScalar}
Simone L~E and Will C~M 1992 {\em Class. Quant. Grav.\/} {\bf 9} 963--978

\bibitem{Ohashi2004_MassiveScalarQ}
Ohashi A and Sakagami M~a 2004 {\em Class. Quant. Grav.\/} {\bf 21} 3973--3984
  (\textit{Preprint} \eprint{gr-qc/0407009})

\bibitem{Konoplya2004_MassiveScalar}
Konoplya R~A and Zhidenko A~V 2005 {\em Phys. Lett. B\/} {\bf 609} 377--384
  (\textit{Preprint} \eprint{gr-qc/0411059})

\bibitem{Dolan2007_MassiveScalarKerr}
Dolan S~R 2007 {\em Phys. Rev. D\/} {\bf 76} 084001 (\textit{Preprint}
  \eprint{0705.2880})

\bibitem{refDecanini2011}
Decanini Y, Folacci A and Raffaelli B 2011 {\em Phys. Rev. D\/} {\bf 84} 084035
  (\textit{Preprint} \eprint{1108.5076})

\bibitem{Konoplya2019_recipes}
Konoplya R~A, Zhidenko A and Zinhailo A~F 2019 {\em Class. Quant. Grav.\/} {\bf
  36} 155002 (\textit{Preprint} \eprint{1904.10333})

\bibitem{Lagos2020_Anomalous}
Lagos M, Ferreira P~G and Tattersall O~J 2020 {\em Phys. Rev. D\/} {\bf 101}
  084018 (\textit{Preprint} \eprint{2002.01897})

\bibitem{Fontana2020_RNanomalous}
Fontana R~D~B, Gonz{\'a}lez P~A, Papantonopoulos E and V{\'a}squez Y 2021 {\em
  Phys. Rev. D\/} {\bf 103} 064005 (\textit{Preprint} \eprint{2011.10620})

\bibitem{Papantonopoulos2022}
Gonz{\'a}lez P~A, Papantonopoulos E, Saavedra J and V{\'a}squez Y 2022 {\em
  JHEP\/} {\bf 06} 150 (\textit{Preprint} \eprint{2204.01570})

\bibitem{GibbonsHawking1977_BHTherm}
Gibbons G~W and Hawking S~W 1977 {\em Phys. Rev. D\/} {\bf 15} 2738--2751

\bibitem{MossNaritaka2021_KerrdSevaporation}
Gregory R, Moss I~G, Oshita N and Patrick S 2021 {\em Class. Quant. Grav.\/}
  {\bf 38} 185005 (\textit{Preprint} \eprint{2103.09862})

\bibitem{ArkaniHamed2006_WGCorigins}
{Arkani-Hamed} N, Motl L, Nicolis A and Vafa C 2007 {\em JHEP\/} {\bf 06} 060
  (\textit{Preprint} \eprint{hep-th/0601001})

\bibitem{Lee:2021cor}
Lee S~M, Cheong D~Y, Hyun S~C, Park S~C and Seo M~S 2022 {\em JHEP\/} {\bf 02}
  100 (\textit{Preprint} \eprint{2111.04010})

\bibitem{SCParkDYCheong2022_FLmili}
Ban K, Cheong D~Y, Okada H, Otsuka H, Park J~C and Park S~C 2023 {\em PTEP\/}
  {\bf 2023} 013B04

\bibitem{Guidetti:2022AFL}
Guidetti V, Righi N, Venken V and Westphal A 2023 {\em JHEP\/} {\bf 01} 114
  (\textit{Preprint} \eprint{2206.03494})

\bibitem{Romans1991_ColdLukewarmRNdS}
Romans L~J 1992 {\em Nucl. Phys. B\/} {\bf 383} 395--415 (\textit{Preprint}
  \eprint{hep-th/9203018})

\bibitem{Belgiorno2009_chargedBHs}
Belgiorno F, Cacciatori S~L and Dalla~Piazza F 2009 {\em JHEP\/} {\bf 08} 028
  (\textit{Preprint} \eprint{0906.1520})

\bibitem{Belgiorno2010_chargedBHs}
Belgiorno F, Cacciatori S~L and Dalla~Piazza F 2010 {\em Class. Quant. Grav.\/}
  {\bf 27} 055011 (\textit{Preprint} \eprint{0909.1454})

\bibitem{AntoniadisBenakli2020_WGCdS}
Antoniadis I and Benakli K 2020 {\em Fortschritte der Physik\/} {\bf 68}
  2000054 (\textit{Preprint} \eprint{2006.12512})

\bibitem{Bekenstein2003_BHinfo}
Bekenstein J~D 2003 {\em Contemp. Phys.\/} {\bf 45} 31--43 (\textit{Preprint}
  \eprint{quant-ph/0311049})

\bibitem{DiasReallSantos2018_SCC}
Dias O~J~C, Reall H~S and Santos J~E 2019 {\em Class. Quant. Grav.\/} {\bf 36}
  045005 (\textit{Preprint} \eprint{1808.04832})

\bibitem{Hawking1974_HawkT}
Hawking S~W 1974 {\em Nature\/} {\bf 248} 30--31

\bibitem{MossMyers1998_CC}
Brady P~R, Moss I~G and Myers R~C 1998 {\em Phys. Rev. Lett.\/} {\bf 80}
  3432--3435 (\textit{Preprint} \eprint{gr-qc/9801032})

\bibitem{Chrysostomou:2024inc}
Chrysostomou A, Cornell A, Deandrea A, Noshad H and Park S~C 2025 {\em PoS\/}
  {\bf ICHEP2024} 782

\bibitem{Brill1993_RNdSextrema}
Brill D~R and Hayward S~A 1994 {\em Class. Quant. Grav.\/} {\bf 11} 359--370
  (\textit{Preprint} \eprint{gr-qc/9304007})

\bibitem{Mann1995_ChargedBHpairs}
Mann R~B and Ross S~F 1995 {\em Phys. Rev. D\/} {\bf 52} 2254--2265
  (\textit{Preprint} \eprint{gr-qc/9504015})

\bibitem{Bousso1999_QuantumStructredS}
Bousso R 1999 {\em Phys. Rev. D\/} {\bf 60} 063503 (\textit{Preprint}
  \eprint{hep-th/9902183})

\bibitem{refNatarioSchiappa}
Nat{\'a}rio J and Schiappa R 2004 {\em Adv. Theor. Math. Phys.\/} {\bf 8}
  1001--1131 (\textit{Preprint} \eprint{hep-th/0411267})

\bibitem{refIKrn}
Kodama H and Ishibashi A 2004 {\em Prog. Theor. Phys.\/} {\bf 111} 29--73
  (\textit{Preprint} \eprint{hep-th/0308128})

\bibitem{HawkingEllis1973_LSSUbook}
Hawking S~W and Ellis G~F~R 1973 {\em The Large Scale Structure of
  Space-Time\/} Cambridge Monographs on Mathematical Physics (Cambridge:
  Cambridge University Press)

\bibitem{BritoCardosoPani2020_Superradiance}
Brito R, Cardoso V and Pani P 2020 {\em Superradiance\/} (Springer
  International Publishing) (\textit{Preprint} \eprint{1501.06570})

\bibitem{KonoplyaZhidenko2014_RNdSrprc}
Konoplya R~A and Zhidenko A 2014 {\em Phys. Rev. D\/} {\bf 90} 064048
  (\textit{Preprint} \eprint{1406.0019})

\bibitem{Hod2018_RNdS}
Hod S 2018 {\em Phys. Lett. B\/} {\bf 786} 217 [Erratum: Phys.Lett.B 796, 256
  (2019)] (\textit{Preprint} \eprint{1808.04077})

\bibitem{DiasSantos2020_RNdSinstability}
Dias O~J~C and Santos J~E 2020 {\em Phys. Rev. D\/} {\bf 102} 124039
  (\textit{Preprint} \eprint{2005.03673})

\bibitem{refNollert1999}
Nollert H~P 1999 {\em Class. Quant. Grav.\/} {\bf 16} R159--R216

\bibitem{refBHWKB0}
Schutz B~F and Will C~M 1985 {\em Astrophys. J.\/} {\bf 291} L33 ISSN 0004-637X

\bibitem{refBHWKB0.5}
Will C~M 1986 {\em Can. J. Phys.\/} {\bf 64} 140--145 ISSN 0008-4204

\bibitem{refBHWKB1}
Iyer S and Will C~M 1987 {\em Phys. Rev. D\/} {\bf 35} 3621--3631 ISSN 05562821

\bibitem{Chrysostomou2023_EPJC}
Chrysostomou A, Cornell A, Deandrea A, Ligout {\'E} and Tsimpis D 2023 {\em
  Eur. Phys. J. C\/} {\bf 83} 325 (\textit{Preprint} \eprint{2211.08489})

\bibitem{ZhuZhang2014_RNdSinstability}
Zhu Z, Zhang S~J, Pellicer C~E, Wang B and Abdalla E 2014 {\em Phys. Rev. D\/}
  {\bf 90} 044042 [Addendum: Phys.Rev.D 90, 049904 (2014)] (\textit{Preprint}
  \eprint{1405.4931})

\bibitem{Furuhashi:2004jk}
Furuhashi H and Nambu Y 2004 {\em Prog. Theor. Phys.\/} {\bf 112} 983--995
  (\textit{Preprint} \eprint{gr-qc/0402037})

\bibitem{KonoplyaZhidenko2013_dRNdSinstability}
Konoplya R~A and Zhidenko A 2014 {\em Phys. Rev. D\/} {\bf 89} 024011
  (\textit{Preprint} \eprint{1309.7667})

\bibitem{DiasEperonReallSantos2018_SCC}
Dias O~J~C, Eperon F~C, Reall H~S and Santos J~E 2018 {\em Phys. Rev. D\/} {\bf
  97} 104060 (\textit{Preprint} \eprint{1801.09694})

\bibitem{Konoplya2024_TwoRegimes}
Konoplya R~A 2024 {\em Phys. Rev. D\/} {\bf 109} 104018 (\textit{Preprint}
  \eprint{2401.17106})

\bibitem{Vishveshwara1970_stability}
Vishveshwara C~V 1970 {\em Phys. Rev. D\/} {\bf 1} 2870--2879

\bibitem{Hatsuda2019_WKB}
Hatsuda Y 2020 {\em Phys. Rev. D\/} {\bf 101} 024008 (\textit{Preprint}
  \eprint{1906.07232})

\bibitem{BenderWu1969_AnharmonicOscillator}
Bender C~M and Wu T~T 1969 {\em Phys. Rev.\/} {\bf 184} 1231--1260

\bibitem{PoschlTellerMethod}
Blome H~J and Mashhoon B 1984 {\em Physics Letters A\/} {\bf 100} 231--234 ISSN
  0375-9601

\bibitem{refFerrMashh1}
Ferrari V and Mashhoon B 1984 {\em Phys. Rev. D\/} {\bf 30} 295--304

\bibitem{refFerrMashh2}
Ferrari V and Mashhoon B 1984 {\em Phys. Rev. Lett.\/} {\bf 52} 1361

\bibitem{Dias2009_Paraspectral1}
Dias O~J~C, Figueras P, Monteiro R, Santos J~E and Emparan R 2009 {\em Phys.
  Rev. D\/} {\bf 80} 111701

\bibitem{Dias2009_Paraspectral2}
Dias O~J~C, Figueras P, Monteiro R, Reall H~S and Santos J~E 2010 {\em JHEP\/}
  {\bf 05} 076

\bibitem{Cardoso2017_QNMsSCC}
Cardoso V, Costa J~L, Destounis K, Hintz P and Jansen A 2018 {\em Phys. Rev.
  Lett.\/} {\bf 120} 031103 (\textit{Preprint} \eprint{1711.10502})

\bibitem{Mo2018_SCC}
Mo Y, Tian Y, Wang B, Zhang H and Zhong Z 2018 {\em Phys. Rev. D\/} {\bf 98}
  124025 (\textit{Preprint} \eprint{1808.03635})

\bibitem{Cardoso2018_QNMsSCC}
Cardoso V, Costa J~L, Destounis K, Hintz P and Jansen A 2018 {\em Phys. Rev.
  D\/} {\bf 98} 104007 (\textit{Preprint} \eprint{1808.03631})

\bibitem{BertiBrito2019_UltraLight1GW}
Hannuksela O~A, Wong K~W~K, Brito R, Berti E and Li T~G~F 2019 {\em Nature
  Astron.\/} {\bf 3} 447--451 (\textit{Preprint} \eprint{1804.09659})

\bibitem{BertiBrito2017a_UltraLightStochastic}
Brito R, Ghosh S, Barausse E, Berti E, Cardoso V, Dvorkin I, Klein A and Pani P
  2017 {\em Phys. Rev. Lett.\/} {\bf 119} 131101 (\textit{Preprint}
  \eprint{1706.05097})

\bibitem{BertiBrito2017b_UltraLightLIGOLISA}
Brito R, Ghosh S, Barausse E, Berti E, Cardoso V, Dvorkin I, Klein A and Pani P
  2017 {\em Phys. Rev. D\/} {\bf 96} 064050 (\textit{Preprint}
  \eprint{1706.06311})

\bibitem{Nariai1950_StaticSol}
Nariai H 1950 {\em Sci. Rep. Tohoku Univ. Eighth Ser.\/} {\bf 34} 160

\bibitem{Nariai1951_NewSol}
Nariai H 1951 {\em Sci. Rep. Tohoku Univ. Eighth Ser.\/} {\bf 35} 46

\bibitem{Cardoso2003_ExtremalSchw}
Cardoso V and Lemos J~P~S 2003 {\em Phys. Rev. D\/} {\bf 67} 084020
  (\textit{Preprint} \eprint{gr-qc/0301078})

\bibitem{refCardosoDiasLemos2004_Nariai-BR-antiNariai}
Cardoso V, Dias O~J~C and Lemos J~P~S 2004 {\em Phys. Rev. D\/} {\bf 70} 024002
  (\textit{Preprint} \eprint{hep-th/0401192})

\bibitem{CasalsDolan2009_Nariai}
Casals M, Dolan S~R, Ottewill A~C and Wardell B 2009 {\em Phys. Rev. D\/} {\bf
  79} 124043 (\textit{Preprint} \eprint{0903.0395})

\bibitem{GinspargPerry1982_NariaidS}
Ginsparg P~H and Perry M~J 1983 {\em Nucl. Phys. B\/} {\bf 222} 245--268

\bibitem{HawkingRoss1995_NariaiRNdS}
Hawking S~W and Ross S~F 1995 {\em Phys. Rev. D\/} {\bf 52} 5865--5876
  (\textit{Preprint} \eprint{hep-th/9504019})

\bibitem{Bertotti1959_BRsol}
Bertotti B 1959 {\em Phys. Rev.\/} {\bf 116} 1331

\bibitem{Robinson1959_BRsol}
Robinson I 1959 {\em Bull. Acad. Pol. Sci. Ser. Sci. Math. Astron. Phys.\/}
  {\bf 7} 351--352

\bibitem{Anninos2012_Musings}
Anninos D 2012 {\em Int. J. Mod. Phys. A\/} {\bf 27} 1230013 (\textit{Preprint}
  \eprint{1205.3855})

\bibitem{Konoplya2003}
Konoplya R~A 2003 {\em Phys. Rev. D\/} {\bf 68} 024018 (\textit{Preprint}
  \eprint{gr-qc/0303052})

\bibitem{Dias2015_Num}
Dias {\'O}~J~C, Santos J~E and Way B 2016 {\em Class. Quant. Grav.\/} {\bf 33}
  133001 (\textit{Preprint} \eprint{1510.02804})

\bibitem{refGoebel1972}
Goebel C~J 1972 {\em Astrophys. J.\/} {\bf 172}

\end{thebibliography}
\end{document}